\documentclass[aps,prb,twocolumn,superscriptaddress]{revtex4}
\usepackage{amsmath,amssymb,graphicx,bm}

\begin{document}

\title{Conductance through a potential barrier embedded in a Luttinger
liquid: \\ nonuniversal scaling at strong coupling. }
\author{D.N. Aristov}
\altaffiliation[On leave from ]{Petersburg Nuclear Physics Institute, Gatchina 188300, Russia}
\affiliation{Institut f\"{u}r Theorie der Kondensierten Materie and Center for Functional
Nanostructures, Universit\"{a}t Karlsruhe, 76128 Karlsruhe, Germany}
\author{P. W\"olfle}
\affiliation{Institut f\"{u}r Theorie der Kondensierten Materie and Center for Functional
Nanostructures, Universit\"{a}t Karlsruhe, 76128 Karlsruhe, Germany}
\affiliation{Institut f\"ur Nanotechnologie, Forschungszentrum Karlsruhe, 76021
Karlsruhe, Germany}
\date{\today}

\begin{abstract}
We calculate the linear response conductance of electrons in a Luttinger
liquid with arbitrary interaction $g_{2}$, and subject to a potential
barrier of arbitrary strength, as a function of temperature. We map the
Hamiltonian in the basis of scattering states into an effective low energy
Hamiltonian in current algebra form. First the renormalization group (RG)
equation for weak interaction is derived in the current operator language
both using the operator product expansion and the equation of motion method.
To access the strong coupling regime, two methods of deducing the RG
equation from perturbation theory, based on the scaling hypothesis and on
the Callan-Symanzik formulation, are discussed. The important role of scale
independent terms is emphasized. The latter depend on the regulaization
scheme used (length versus temperature cutoff). Analyzing the perturbation
theory in the fermionic representation, the diagrams contributing to the
renormalization group $\beta $-function are identified. A universal part of
the $\beta $-function is given by a ladder series and summed to all orders
in $g_{2}$. First non-universal corrections beyond the ladder series are
discussed and are shown to differ from the exact solutions obtained within
conformal field theory which use a different regularization scheme. The RG
equation for the temperature dependent conductance is solved analytically.
Our result agrees with known limiting cases.
\end{abstract}

\maketitle

\section{Introduction}

The quantum theory of charge transport in one-dimensional (1D) systems of
interacting fermions is a fundamental building block of our understanding of
electron transport in nanostructures and of a future nanoelectronics.
One-dimensional interacting fermion systems have been studied since the
1950s, beginning with the pioneering works of Tomonaga and Luttinger. The
early work exploited the fact that the elementary excitations in these
systems are charge and spin excitations of bosonic character, and led to the
formulation of the method of bosonization. \cite{GoNeTs,GiamarchiBook} This
method offers a natural description of the fractionalization of the usual
fermionic quasiparticles existing in higher dimensions, into spinon and
holon quasiparticles in 1D. While the bosonization method is highly
successful in accounting for the properties of systems in equilibrium, it is
somewhat problematic when applied to transport situations. For example, an
ideal quantum wire attached to two reservoirs is expected to have a
two-point conductance of $R_{Q}=\frac{e^{2}}{h}$ per channel. In
contradiction to this general law it was found in early work on transport in
clean Luttinger liquids \cite{Apel1982, Kane1992,Kane1992a} that the
conductance (here and in the following conductance will be measured in units
of the conductance quantum $R_{Q}$) is given by the Luttinger parameter $K$,
and thus depends on the interaction strength. It was later pointed out that
by carefully choosing the order of limits of frequency, $\omega \rightarrow
0 $, and wire length, $L\rightarrow \infty $, \cite{Maslov1995,Safi1995} \
or else by taking into account the screening of the external field by the
interacting electron system \cite{Oreg1996} \ one recovers the correct value
of unity for the two-point conductance. Whether the former or the latter
explanation is the correct one remains disputed to this day.

The more relevant case of a Luttinger liquid with potential barrier has been
considered first by Kane and Fisher \cite{Kane1992,Kane1992a} and by
Furusaki and Nagaosa, \cite{Furusaki1993} again using the bosonization
method. These authors found that interaction has a dramatic effect on the
conductance: for repulsive interaction the conductance is found to tend to
zero as a fractional power of the temperature $T$ , in the limit $%
T\rightarrow 0$ . This was shown in the two limits of a weakly scattering
barrier and a strong (tunneling) barrier. In addition, results in the
complete temperature range are available at special values of the
interaction parameter, $K=\frac{1}{2}$ and $K=\frac{1}{3}$. \cite%
{Kane1992,Kane1992a,Weiss1995,Fendley1995,Fendley1995a} All these works
suffer from the drawback that in the clean limit the conductance is found to
tend to $K$ rather than $1$ . It has been argued that these results may be
applied to the four-point conductance. However, the four-point conductance
is expected to tend to infinity in the limit of vanishing barrier strength,
a behavior not shown.

For the reasons noted above we think it worthwhile to develop an alternative
formulation of the transport theory of Luttinger liquids, formulated
entirely in the fermionic language. The fermionic representation offers the
possibility to connect the fermionic degrees of freedom in the
(non-interacting) leads smoothly with the degrees of freedom in the
interacting system. In other words, it allows to use the scattering states
of the system in the non-interacting limit as a basis of description. On the
simplest level, in lowest order in the interaction, this program has been
carried out in a seminal paper by Yue, Matveev and Glazman. \cite{Yue1994}
The physics of this problem lies in the scale-dependent build-up of a
polarization potential around the bare barrier, induced by the Friedel
oscillations of the density. For repulsive interaction, the polarization
potential is found to extend farther and farther out as the temperature is
lowered, until at $T=0$ it is infinitely extended, leading to a vanishing
transmission probability across the barrier. The process of gradual growth
of the effective potential barrier may be described in the language of
renormalization group theory applied to the transmission probability as a
function of the temperature. A generalization of the approach of Yue et al.
to the case of two barriers has been given in Ref.\ [%
\onlinecite{Polyakov2003}].  These predictions, based on the continuum 
version of the theory, were thoroughly checked and confirmed by the 
fermionic functional renormalization group method, starting from 
the Hubbard models on a lattice. \cite{Enss2005,Andergassen2006}

In this paper we report on a significant extension of the work of Yue et al.
to the case of systems of spinless fermions in 1d interacting via
arbitrarily strong forward scattering (parameter $g_{2}$) and subject to a
short range barrier potential (width $a$). The principal tool we will be
using is perturbation theory in the interaction, summed to infinite order.
To achieve this goal, and to gain insight into the structure of perturbation
theory, we map the problem first onto an effective model of interacting
currents of chiral fermions. The reformulation allows to describe the effect
of the barrier potential as a local magnetic field at the position of the
barrier, acting on the pseudospin vector of the chiral current. The
logarithmic corrections to this field, stemming from the fermionic
interactions can be addressed in a simplified poor-man scaling approach;
however this approach becomes ambiguous in higher orders. Therefore at a
later stage we will apply the Callan-Szymanzik type of analysis, by first
introducing the counter terms to magnetic field part of the Hamiltonian to
compensate the effect of the interaction, and secondly requiring the
``bare'' value of the field to be independent of the high-energy cutoff.
Equivalently, we may assume the scaling property of the conductance to hold,
entailing the existence of a renormalization group equation for the
conductance as a function of the scaling variable $\Lambda $ (at finite
temperature $\Lambda =\ln (T_{0}/T)$ ). By comparison with the structure of
perturbation theory for the conductance, which is given by a power series in
the interaction and in the scaling parameter $\Lambda $ , one may extract
the RG $\beta $-function. In this way we derive a universal renormalization
group equation for the conductance as a function of the scaling parameter.

The most important consequence of the current algebra formulation is,
however, to allow for an efficient organization of perturbation theory.
After presenting the formulation of perturbation theory as well as a few low
order results we analyze the structure of the perturbation expansion with
respect to logarithmically divergent terms containing powers of \ $\Lambda
=\ln (L/a)$ \ or \ $\ln (T_{0}/T)$. We identify a class of universal
diagrams contributing the principal terms linear in $\Lambda $ and derive an
integral equation resumming these contributions (``ladder approximation'').
These terms are shown to dominate in both limits, small and large
conductance. We identify the prefactor of the terms in the perturbation
series for the conductance $G$ linear in $\Lambda $ with the renormalization
group $\beta $-function of the flow equation of $G(\Lambda )$. There is,
however, one important new feature: there appear scale independent terms
(from third order on), which lead to a correction of the conductance even at 
$\Lambda =0$, the ultraviolet cutoff. These terms must be taken into
account, in order to preserve the scaling property of the conductance, and
lead to a starting value of the conductance $G(\Lambda =0)$, different from
the conductance in the absence of interaction.

The RG equation may be solved analytically in the ladder approximation, to
give the conductance as a function of temperature for any interaction
strength and barrier potential. The result completely agrees with earlier
work, where a comparison is appropriate We emphasize that our result for the
conductance reduces to unity in the absence of a barrier. Furthermore, the
effect of screening of the external field \cite{Oreg1996}\ \ is included by
omitting all polarization diagrams. We checked the renormalizability of the
theory explicitly by calculating all terms up to third order in $\Lambda $
and $g_{2}$ .

At intermediate temperatures (semi-transparent barrier) additional diagrams
contribute small corrections to the $\beta $-function. We calculate these
terms in lowest (third) order in $g_{2}$ . A proper treatment of these
contributions requires taking into account scale independent terms in a way
sketched above. We demonstrate explicitly that the resulting $\beta $%
-function is uniquely determined for a given physical quantity. However, $%
\beta $ turns out to be a slightly different function for the two cases of
interest here, the temperature dependent and the length dependent
conductance. The result of the solution of the RG-equation for the
conductance, using the approximate $\beta $-function thus obtained, is found
to agree quantitatively with all known results on the scaling behavior of
the conductance in the limits $G\rightarrow 0$ and $G\rightarrow 1$ . In the
intermediate regime $G\sim 1/2$ we find small discrepancies with the results
of conformal field theory in the case $K=\frac{1}{2}$, which we can trace
back to a different cutoff scheme used there and the effect of higher order
terms neglected in our work. As an excellent interpolation formula between
two exact limits our result goes beyond all previous works: it is valid for
any interaction strength $K$ and any potential scattering strength.

\subsection{Formulation of the problem}

The Hamiltonian includes the kinetic energy of freely right- and left-moving
(R,L) spinless fermions with linearized dispersion, $H_{0}$, the interaction
energy between the left- and right-moving densities, $H_{1}$, and the
impurity part decribing scattering off the barrier, placed at the origin, $%
H_{imp}$ : 
\begin{eqnarray}
H &=&H_{0}+H_{1}+H_{imp}  \label{genHam} \\
H_{0} &=&v_{F}\int_{-\infty }^{\infty }{dx}\left[ \psi _{R}^{\dagger
}(i\partial _{x})\psi _{R}-\psi _{L}^{\dagger }(i\partial _{x})\psi _{L}%
\right]  \notag \\
H_{1} &=&g_{2}\int_{-L}^{L}{dx}\;(\psi _{R}^{\dagger }\psi _{R})(\psi
_{L}^{\dagger }\psi _{L})  \notag
\end{eqnarray}%
In order to regularize the theory, we assume that the region of interaction $%
|x|>a$, where $g_{2}\neq 0$, and the region over which the impurity
potential is nonzero, $|x|<a\sim k_{F}$ ($k_{F}$is the Fermi wave vector) do
not overlap spatially. The length scale $a$ will serve as an ultraviolet
cutoff in the scale dependent logarithmic corrections considered below.

The last term, $H_{imp},$ in (\ref{genHam}) is not explicitly specified here
(but see below). We will rather take the one-particle scattering states to
be known. This means that we characterize the barrier by transmission and
reflection amplitudes $t=\tilde{t}=\cos \theta $, $r=-\tilde{r}^{\ast
}=i\sin \theta e^{i\phi }$ , with negligible energy dependence in the energy
range of interest (here $t$, $\tilde{t}$, $r$, $\tilde{r}$ are the usual
components of the impurity scattering matrix). Notice also that we assume
the system to be non-interacting in the leads, $|x|>L$, which allows for an
asymptotic single-particle scattering states representation.

We define the creation operator of scattering states as 
\begin{eqnarray}
\psi _{k}^{\dag }(x) &=&(e^{ikx}+r_{k}e^{-ikx})c_{1k}^{\dagger }+\tilde{t}%
_{k}e^{-ikx}c_{2k}^{\dagger },\quad x<0  \notag \\
&=&t_{k}e^{ikx}c_{1k}^{\dagger }+(\tilde{r}_{k}e^{ikx}+e^{-ikx})c_{2k}^{%
\dagger },\quad x>0  \label{scat-states}
\end{eqnarray}%
with $c_{1k}^{\dagger }$ ($c_{2k}^{\dagger }$) the creation operators of
asymptotically right-going (left-going) fermions of momentum $k$. Such a
representation requires the kinetic energy part of the Hamiltonian to be of
the form $-\nabla ^{2}/(2m)$, and the momentum, $k=\sqrt{2mE}>0$, before the
linearization of the dispersion around the two Fermi points. In appendix A
we clarify the correspondence between the scattering states representation (%
\ref{scat-states}) and our linearized Hamiltonian (\ref{genHam}) written in
the basis of chiral fermions.

For simplicity we do not consider the so-called $g_{4}$ part of the
fermionic interaction in this work, i.e. the term $\frac{1}{2}%
g_{4}\int_{-L}^{L}{dx}\;[(\psi _{R}^{\dagger }\psi _{R})^{2}+(\psi
_{L}^{\dagger }\psi _{L})^{2}]$. For $L\rightarrow \infty $ the $g_{4}$%
-interaction can be absorbed into the redefinition of the Fermi velocity, $%
\tilde{v}_{F}=v_{F}+g_{4}/2\pi $. For finite $L$ one can show that $g_{4}$
does not lead to a renormalization of d.c.\ conductance, which is the
quantity of interest below. The effect of $g_{4}$ on the a.c. conductance
can be analyzed, following the guidelines in \cite{Oreg1995,Safi1995}.

%\subsection{Impurity potential and the basis of scattering states}
\label{sec:ScatteringStates}

It appears that in the case of weak potential scattering, $U(x)\ll E_{F}$,
and for electrons with energies close to the Fermi energy, $E_{F}$, the
situation is characterized two limits of the Fourier transform $U(q)$ of the
potential: the forward scattering amplitude, $U(q\simeq 0)=u_{1}$, and the
backward scattering amplitude, $U(q\simeq 2k_{F})=u_{2}$, with real-valued $%
u_{1}$ and complex-valued $u_{2}$. The fermionic Hamiltonian $H_{imp}$ in (%
\ref{genHam}) is then written as 
\begin{eqnarray}
H_{imp} &=&v_{F}\int {dx}\left[ u_{1}(x)(\psi _{R}^{\dagger }\psi _{R}+\psi
_{L}^{\dagger }\psi _{L})\right.  \label{impHam} \\
&&\left. +(u_{2}(x)\psi _{L}^{\dagger }\psi _{R}+h.c.)\right] ,  \notag
\end{eqnarray}%
where the dimensionless functions $u_{1,2}(x)$ are short-range impurity
potentials, which means that $u_{1,2}(x)=0$ for $|x|>a$ and the above
amplitudes $u_{1,2}=\int dx\,u_{1,2}(x)$.

In Appendix \ref{sec:appSmatrix} we show that the connection between the
microscopic Hamiltonian and the $S$-matrix can be clarified in some simple
cases, but requires a detailed knowledge of $u_{1,2}(x)$, when one goes
beyond the lowest Born approximation.

Returning to (\ref{scat-states}), we define Fourier transforms, $\psi
_{1}^{+}(x)=\int_{0}^{\infty }\frac{dk}{2\pi }e^{ikx}\tilde{c}_{1,k}^{+}$
and $\psi _{2}^{+}(x)=\int_{0}^{\infty }\frac{dk}{2\pi }e^{-ikx}\tilde{c}%
_{2,k}^{+}$. Then we have for the electron creation operator at position $x$ 
\begin{eqnarray}
\psi ^{+}(x) &=&\left\{ \Theta (-x)\left[ \psi _{1}^{+}(x)+r\psi
_{1}^{+}(-x)+\tilde{t}\psi _{2}^{+}(x)\right] \right.  \notag \\
&+&\left. \Theta (x)\left[ t\psi _{1}^{+}(x)+\tilde{r}\psi _{2}^{+}(-x)+\psi
_{2}^{+}(x)\right] \right\} \,,  \label{ChiralScatStates}
\end{eqnarray}%
with the step function $\Theta (x)=1$ at $x>0$.

%%%%%%%%%%%%%%%%%%%%%%%%%%%%%%%%%%%%%

\section{Interacting case, previous works}

\subsection{Boundary sine-Gordon model}

\label{sec:BSG}

In bosonization technique, the chiral fermions are represented as
exponentials of chiral boson fields, $\psi _{R}^{\dagger }\sim e^{-i\varphi
_{R}}$, $\psi _{L}^{\dagger }\sim e^{i\varphi _{L}}$. with the fields $%
\varphi _{R(L)}=\varphi \mp \theta $. Here the primary field $\varphi $ and
its canonically conjugate momentum $\Pi =\pi ^{-1}\partial _{x}\theta $ obey
the commutation relation $[\varphi (x),\Pi (y)]=i\delta (x-y)$. We have for
the density $\psi _{R}^{\dagger }\psi _{R}+\psi _{L}^{\dagger }\psi
_{L}=\partial _{x}\varphi /\pi $ and the Hamiltonian (\ref{genHam}), (\ref%
{impHam}) can be represented as 
\begin{eqnarray}
H &=&\frac{\tilde{v}_{F}}{2\pi }\int dx\Big[K\,(\partial _{x}\theta
)^{2}+K^{-1}(\partial _{x}\varphi )^{2}  \notag \\
&&+2u_{1}(x)\partial \varphi (x)+u_{2}(x)\cos 2\varphi (x)\Big]
\label{bosonHam}
\end{eqnarray}%
One has $\tilde{v}_{F}=v_{F}$ and $K=1$ for free fermions, whereas the
short-range interaction $H_{1}$ between the fermions, Eq.\ (\ref{genHam}),
leads to the renormalized Fermi velocity $\tilde{v}_{F}=v_{F}(1-g^{2})^{1/2}$
and the Luttinger parameter $K=[(1-g)/(1+g)]^{1/2}$. Here and below we use 
\begin{equation}
g=g_{2}/(2\pi v_{F})
\end{equation}

One usually argues, that the term $u_{1}$ can be absorbed into a
redefinition of $\varphi (x)\rightarrow \varphi (x)+u_{1}sgn(x)$ (cf.
Appendix \ref{sec:appSmatrix}). After this redefinition one concentrates on
the $u_{2}$ term which constitutes the boundary sine-Gordon (BSG) problem. A
solution to this problem, in the lowest order of $u_{2}$ was presented in
the early work. \cite{Kane1992,Kane1992a} Focussing on the small-energy
sector of the problem, one uses the renormalization group approach, removing
the higher energy states of the problem while simultaneously rescaling the
parameters of the action. It turns out that under this procedure the
amplitude of the BSG term is renormalized as 
\begin{equation}
u_{2}\rightarrow u_{2}e^{(1-K)\Lambda }
\end{equation}%
where the RG scaling variable $\Lambda =\ln E_{F}/\epsilon >0$ . As a
result, the conductance at small $u_{2}$ , given by $G=|t|^{2}\simeq
1-u_{2}^{2}$ , undergoes a similar renormalization 
\begin{equation}
G\simeq 1-|u_{2}|^{2}e^{2(1-K)\Lambda }  \label{conduKF.high}
\end{equation}%
with the lower cutoff in $\Lambda $ defined by temperature or voltage bias, $%
\Lambda =\ln E_{F}/\max [T,V]$. We shall confine ourselves to the the linear
response regime, $V\ll T$ \ in the following.

For repulsive interaction $K<1$ the conductance decreases with falling
temperature. When the renormalized impurity potential becomes strong, $%
|u_{2}|^{2}e^{2(1-K)\Lambda }\sim 1$, the weak-impurity assumption is
violated and one should use a different line of reasoning. One may think of
two semi-infinite parts of the wire, connected by a weak tunneling link $t$.
An appropriate RG treatment shows that the repulsive interaction leads to a
further decrease of the conductance, which acquires the form 
\begin{equation}
G\simeq |t|^{2}e^{2(K^{-1}-1)\Lambda }  \label{conduKF.low}
\end{equation}

These two regimes, Eqs. (\ref{conduKF.high}), (\ref{conduKF.low}) show
different scaling exponents, which are approximately equal for small
interaction 
\begin{equation*}
1- K \simeq K^{-1}-1 \simeq g .
\end{equation*}
A smooth crossover between the two asymptotes (\ref{conduKF.high}) and (\ref%
{conduKF.low}) was first explicitly provided for a special case of
interaction strength, $K=1/2$, which is known as Luther-Emery
refermionization point. The full curve for the conductance was obtained \cite%
{Kane1992,Kane1992a,Weiss1995} in the form of a one-parameter scaling
function.

In subsequent works \cite{Fendley1995,Fendley1995a} the BSG model was
analyzed at arbitrary values of Luttinger parameter, $K$. The set of coupled
integral equations was shown to be finite for $1/K=1,2,3\ldots $ and the
solution for $K=1/3$ was derived, in particular. Corresponding scaling
functions for the conductance were obtained for finite $T$ and voltage bias
on the contact $V$ either by numerical solution of the corresponding
integral equation or as a series in powers of the scaling parameter, see
also \cite{Weiss1996}. For practical purposes this form of representation is
not very convenient.

In principle, the statement of a one-parameter scaling function for the
conductance and the knowledge of its actual form provides a solution for the
temperature renormalization of the barrier apparently at arbitrary strength.
However, it was noted in \cite{Fendley1995,Fendley1995a}, that "the
strong-barrier problem which is at the end of our renormalization-group
trajectory follows formally from dimensional continuation of the integrals
for the weak barrier problem. It is not in any case a generic strong-barrier
problem."

\subsection{Fermionic approach, RG for S-matrix}

Yue, Matveev and Glazman \cite{Yue1994} have developed a theory, starting
from the formalism of scattering states and taking into account the
fermionic interaction in lowest order of perturbation. They arrived at the
RG equation for the transmission amplitude $t$ in the form 
\begin{equation}
\frac{dt}{d\Lambda }=-gt(1-|t|^{2})  \label{YMG.RGeq}
\end{equation}

The solution to this equation is found as 
\begin{eqnarray}
t & =& \frac{t_0}{(|t_0|^2 + (1-|t_0|^2) e^{2g\Lambda})^{1/2}}
\label{soluRG1st}
\end{eqnarray}
which, in leading order in $g$, agrees with Eqs. (\ref{conduKF.high}), (\ref%
{conduKF.low}).

The advantage of this approach is its applicability for arbitrary strength
of impurity potential. One may thus avoid the step involving the transition
from the initial Hamiltonian (\ref{bosonHam}) to the observed conductance,
which as we saw above can depend on the short-distance (ultraviolet) details
of relevant quantities. This view is further corroborated by the work by
Callan \textit{et al.} \cite{Callan1994} who discussed the solution of the
non-interacting ($K=1$) BSG theory, and arrived at a current algebra
formulation similar to our formulation below. These authors show that the
connection of the initial model (\ref{bosonHam}) to the observable
quantities depends on the ultraviolet regularization. We will return to this
point in our discussion below.

\section{Scattering states and current algebra}

In the above we defined a scattering states representation (\ref%
{ChiralScatStates}) in chiral representation. In order to describe the
effects of interaction, we need to define fermionic densities in the same
basis. One may form four bilinear density combinations out of two chiral
fermions. It is convenient to group these combinations into chiral densities
or ``currents'', according to

\begin{eqnarray}
\widehat J (x) &=& 
\begin{pmatrix}
\psi^\dagger_1 (x) \psi _1 (x) & \psi^\dagger_1 (x) \psi _2 (-x) \\ 
\psi^\dagger_2 (-x) \psi _1 (x) & \psi^\dagger_2 (-x) \psi _2 (-x)%
\end{pmatrix}
\label{def:currents} \\
&\equiv& 
\begin{pmatrix}
J_0 + J_3 & J_1-iJ_2 \\ 
J_1+iJ_2 & J_0 - J_3%
\end{pmatrix}%
\end{eqnarray}
We call $J_{0}$ the pseudocharge current and the vector $\vec{J}%
=(J_{1},J_{2},J_{3})$ the pseudospin current. These operators obey $U(1)$
and $SU(2)$ Kac-Moody algebras, respectively \cite{GoNeTs,Affleck1991}, as
we discuss below.

The particle density operators for incoming $(i)$ and outgoing $(o)$
particles in terms of the $J_{\mu }$'s are given by (here and in the
following $x>0$ ) 
\begin{eqnarray}
\rho _{iR}(-x) &=&\psi _{1}^{\dagger }(-x)\psi _{1}(-x)=J_{0}(-x)+J_{3}(-x)
\\
\rho _{iL}(x) &=&\psi _{2}^{\dagger }(x)\psi _{2}(x)=J_{0}(-x)-J_{3}(-x) \\
\rho _{oR}(x) &=&(t\psi _{1}^{\dagger }(x)+\tilde{r}\psi _{2}^{\dagger
}(-x))(t^{\ast }\psi _{1}(x)+\tilde{r}^{\ast }\psi _{2}(-x))  \notag \\
&=&(S.\widehat{J}(x).S^{\dagger })_{11}=J_{0}(x)+\widetilde{J}_{3}(x) \\
\rho _{oL}(-x) &=&(r\psi _{1}^{\dagger }(x)+\tilde{t}\psi _{2}^{\dagger
}(-x))(r^{\ast }\psi _{1}(x)+\tilde{t}^{\ast }\psi _{2}(-x))  \notag \\
&=&(S.\widehat{J}(x).S^{\dagger })_{22}=J_{0}(x)-\widetilde{J}_{3}(x)
\end{eqnarray}%
Here $\widetilde{J}_{3}=(R\vec{J})_{3}$ is the third component of the
pseudospin current vector $\vec{J}$ rotated by the orthogonal matrix $R_{\mu
\nu }$ given by 
\begin{equation}
R_{\mu \nu }=\frac{1}{2}Tr[\sigma _{\mu }.S.\sigma _{\nu }.S^{\dagger }].
\label{defR}
\end{equation}%
with Pauli matrices $\sigma _{\mu }$. It is seen from (\ref{defR}) that the
global phase of $S$ drops out completely. Omitting this phase, we
parametrize 
\begin{equation}
S=%
\begin{pmatrix}
t & \tilde{r} \\ 
r & \tilde{t}%
\end{pmatrix}%
=%
\begin{pmatrix}
\cos \theta & i\sin \theta e^{-i\phi } \\ 
i\sin \theta e^{i\phi } & \cos \theta%
\end{pmatrix}
\label{S-parametrization}
\end{equation}%
In this notation, the components of interest below read $%
R_{33}=|t|^{2}-|r|^{2}=\cos 2\theta $, $R_{32}=\mbox{Im}\{t\widetilde{r}%
^{\ast }+\widetilde{t}r^{\ast }\}=-\sin 2\theta \cos \phi $, and $R_{31}=%
\mbox{Re}\{t\widetilde{r}^{\ast }-\widetilde{t}r^{\ast }\}=\sin 2\theta \sin
\phi $.

We see that the description of the barrier is given by the eigenmodes of the
current $J_{i}(x)$, with negative and positive $x$ and $i=0,\ldots ,3$. In
this picture, the incoming current is given by the diagonal components of
the $\widehat{J}$ matrix, and the outgoing currents correspond to the
diagonal components of the rotated matrix $S.\widehat{J}.S^{\dagger }$.
Evidently, the pseudocharge component $J_{0}$ is not affected by this
rotation.

\subsection{Formal identities and Green's functions}

The chiral densities introduced above satisfy the Kac-Moody commutation
rules 
\begin{eqnarray}
\lbrack J_{0}(x),J_{0}(y)] &=&\frac{i}{4\pi }\partial _{x}\delta (x-y)
\label{KacMoody} \\
{}[J_{j}(x),J_{k}(y)] &=&\frac{i}{4\pi }\delta _{jk}\partial _{x}\delta
(x-y)+i\varepsilon _{jkl}J_{l}(y)\delta (x-y)  \notag
\end{eqnarray}%
with $j,k,l=1,\ldots 3$ and totally antisymmetric $\varepsilon _{jkl}$. The
short-distance operator product expansions (OPEs) are 
\begin{eqnarray}
J_{0}(x)J_{0}(y) &=&\frac{-1}{8\pi ^{2}}\frac{1}{(x-y-i0)^{2}}\,,  \notag \\
J_{j}(x)J_{k}(y) &=&\frac{-1}{8\pi ^{2}}\frac{\delta _{jk}}{(x-y-i0)^{2}}+%
\frac{1}{2\pi }\frac{\varepsilon _{jkl}J_{l}(y)}{x-y-i0}\,.  \label{OPE}
\end{eqnarray}%
In simple terms, relevant to our discussion below, one can consider (\ref%
{KacMoody}) as a consequence of (\ref{OPE}). In their turn, the OPEs (\ref%
{OPE}) simply represent fermionic Green's functions, $[2\pi (x-y)]^{-1}$. A
bilinear form in the currents is a product of four fermion operators; the
complete convolution of these gives a product of two two-point Green's
function (modulo normally ordered operators), and the partial convolution
produces a Green's function multiplied by bilinear forms in the fermion
operators, i.e.\ current operators. The position argument of the latter
current operators (last term in (\ref{OPE})) is subject to ambiguity, but
usually this concerns only less relevant terms. In our analysis below we
deal with an action non-local in the currents, and to handle such an
ambiguity becomes an increasingly difficult matter. This is why we resort to
the perturbative treatment in terms of chiral fermions.

\subsection{Observable current and conductance}

The physical charge density is $\rho (x)=\rho _{iR}(x)+\rho _{oL}(x)$ at $%
x<0 $ and $\rho (x)=\rho _{oR}(x)+\rho _{iL}(x)$ at $x>0$. A simple
calculation shows that 
\begin{eqnarray}
\rho (x) &=&\rho _{c}(x)+\rho _{s}(x),  \notag \\
\rho _{c}(x) &=&J_{0}(-x)+J_{0}(x),  \label{def:density} \\
\rho _{s}(x) &=&sign(x)\left( -J_{3}(-|x|)+\widetilde{J}_{3}(|x|)\right) . 
\notag
\end{eqnarray}%
The physical current operator is also decomposed into two parts, in the
pseudocharge and pseudospin sectors. We use the continuity equation $%
-\partial _{x}j(x)=\partial _{t}\rho (x)=-i[\rho (x),H]$ to obtain 
\begin{eqnarray}
j(x) &=&j_{c}(x)+j_{s}(x),  \notag \\
j_{c}(x) &=&v_{F}(-J_{0}(-x)+J_{0}(x)),  \label{def:current} \\
j_{s}(x) &=&v_{F}(J_{3}(-|x|)+\widetilde{J}_{3}(|x|)),  \notag
\end{eqnarray}

It follows then that the pseudocharge current corresponds to the part of
physical density which is even with respect to the reflection at the
scatterer, $\rho _{c}(-x)=\rho _{c}(x)$. The pseudospin current corresponds
to the odd part of the density, $\rho _{s}(-x)=-\rho _{s}(x)$. This symmetry
property simplifies the discussion of a voltage bias, applied to the
barrier. One sees immediately, that applying the same electrochemical
potential to both leads of the wire corresponds to its coupling to the
pseudocharge current. It results in an overall increase of the fermionic
density, symmetric with respect to the barrier, and not to a net physical
current. By contrast, an electric potential of the form $V(x)=\frac{1}{2}%
V\,sign(x)$, couples only to the pseudospin part, leading to a term in the
Hamiltonian 
\begin{equation*}
H_{V}=V\int_{0}^{\infty }dx\,(\widetilde{J}_{3}(x)-J_{3}(-x))
\end{equation*}%
The physical current induced by a bias voltage is therefore given in linear
response theory as 
\begin{eqnarray}
\langle j(x,\omega )\rangle &=&G(x,\omega )V(\omega )  \notag \\
G(x,t) &=&-i\Theta (t)\langle \lbrack j_{s}(x,t),\int_{0}^{\infty }dy\,\rho
_{s}(y,0)]\rangle  \label{def:conduc}
\end{eqnarray}%
A remark is in order here. Strictly speaking, the prefactor $v_{F}$ in the
definition (\ref{def:current}) applies for non-interacting leads. In the
interacting region we have to replace $v_{F}\rightarrow v_{F}-g_{2}/2\pi $
in (\ref{def:current}). We may avoid this complication, analyzed elsewhere 
\cite{Oreg1995, Safi1995}, by assuming in (\ref{def:conduc}) that both $%
j_{s}(x,t)$ and $\rho _{s}(y,0)$ are placed outside the interacting region, $%
x,y>L$. The change of the limits of integration over $y$ in (\ref{def:conduc}%
) is not important in the limit of d.c.\ conductance $|\omega |\ll v_{F}/L$,
considered below.

\begin{figure}[tb]
(a)\includegraphics[width=8cm]{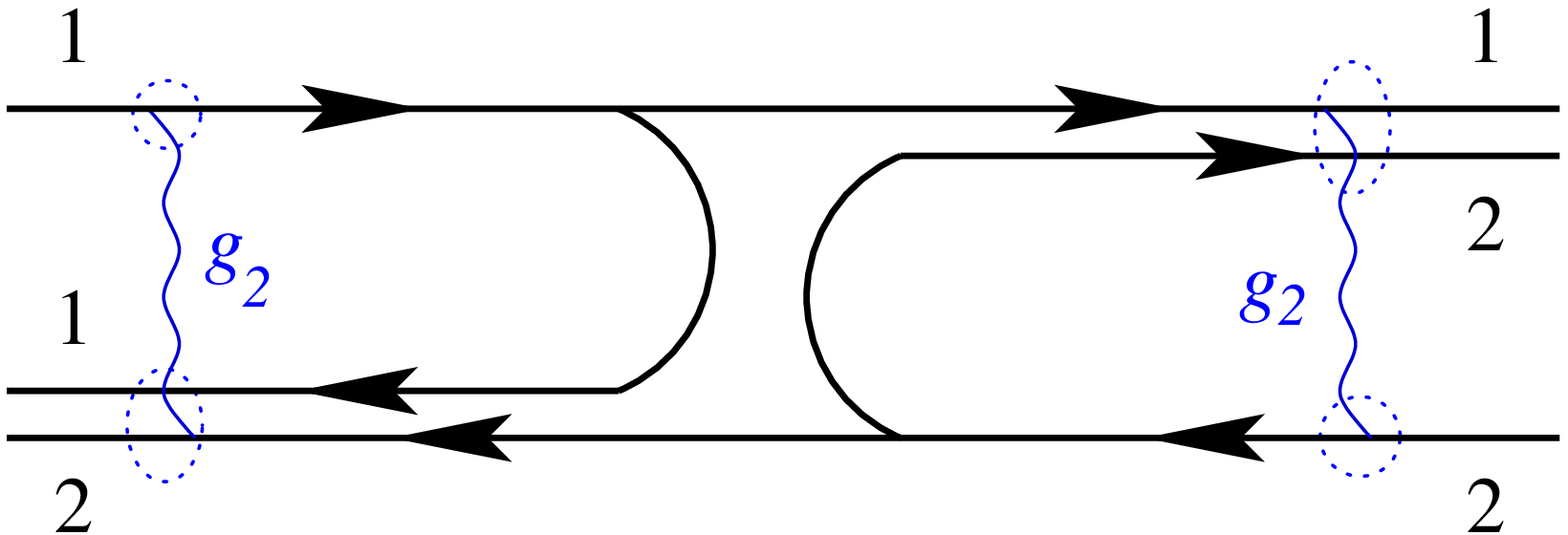} (b)%
\includegraphics[width=8cm]{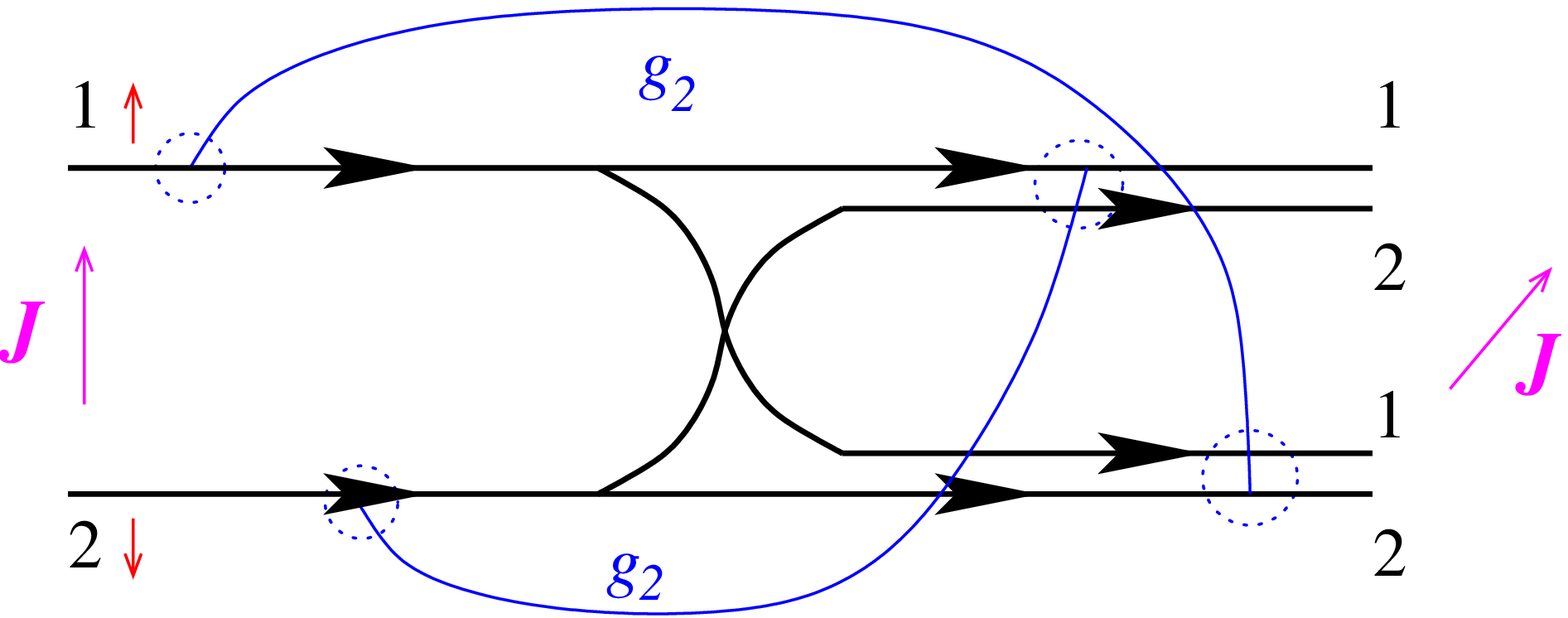}
\caption{(Color online) (a)The interaction of the left- and right-going densities in the
basis of scattered waves. (b) The interaction of the left- and right-going
densities in the chiral basis, corresponding to non-local interaction, Eq.\ (%
\protect\ref{Ham_1}). }
\label{fig:scattering}
\end{figure}

\subsection{Hamiltonian}

It is known that fermions with linearized dispersion in one spatial
dimension can be described by a Hamiltonian, quadratic in chiral fermionic
densities. This observation, usually attributed to Tomonaga, is the basis of
the bosonization approach. \cite{GoNeTs} Similarly to (\ref{bosonHam}), we
have 
\begin{equation*}
H_{0}=\pi v_{F}\int_{0}^{\infty }dx(\rho _{iR}^{2}(-x)+\rho
_{iL}^{2}(x)+\rho _{oR}^{2}(x)+\rho _{oL}^{2}(-x))
\end{equation*}%
which can be represented as 
\begin{eqnarray}
H_{0} &=&2\pi v_{F}\int_{-\infty }^{0}dx\,(J_{0}^{2}(x)+J_{3}^{2}(x))
\label{Ham_0} \\
&&+2\pi v_{F}\int_{0}^{\infty }dx\,(J_{0}^{2}(x)+\widetilde{J}_{3}^{2}(x)) 
\notag
\end{eqnarray}%
The interaction part is 
\begin{eqnarray}
H_{1} &=&g_{2}\int_{a}^{L}dx\,(\rho _{iR}(-x)\rho _{oL}(-x)+\rho
_{oR}(x)\rho _{iL}(x))  \notag \\
&=&2g_{2}\int_{a}^{L}dx\,(J_{0}(-x)J_{0}(x)-J_{3}(-x)\widetilde{J}_{3}(x))
\label{Ham_1}
\end{eqnarray}%
In Fig.\ \ref{fig:scattering} we show in a pictorial way our parametrization
of the fermionic densities and the interaction, $H_{1}$. In Fig.\ \ref%
{fig:scattering}a the $g_{2}$-interaction processes are shown in the usual
scattering configuration. The representation in terms of the chiral currents 
$J_{\mu }$ (all particles moving to the right) is shown in Fig.\ \ref%
{fig:scattering}b and leads to a seemingly nonlocal interaction. Notice that
in the case of perfect reflection, $t=0$, we have $\tilde{J}_{3}=-J_{3}$ and
the observable densities (see below) form an Abelian $U(1)$ sub-algebra of $%
SU(2)$. This case of \textquotedblleft open boundary
bosonization\textquotedblright , considered by Fabrizio and Gogolin, allows
a complete and rather simple analysis \cite{Fabrizio1995} . In the next
subsection we elucidate what happens at the origin, $x=0$. .

\subsection{The barrier as magnetic field}

\label{sec:magfield}

Using the parametrization (\ref{S-parametrization}), one finds that $%
R_{ij}=\exp (\varepsilon _{ijk}B_{k})$ i.e., it is a finite rotation
characterized by the dual vector $\mathbf{B}=2\theta (\cos \phi ,\sin \phi
,0)$. We cannot symmetrize the Hamiltonian in the pseudospin space, reducing
it to the so-called Sugawara form, \cite{GoNeTs} because the vector $\mathbf{%
B}$ breaks the $SU(2)$ symmetry of the problem. We may symmetrize the
Hamiltonian in the plane perpendicular to $\mathbf{B}$, but that provides
little advantage .

The above formulation of the problem, incorporating the knowledge about the
scatterer in the asymptotic scattering states, is equivalent to the the
alternative formulation of the pseudospin current being rotated by a
fictitious magnetic field $\mathbf{B}$ located at the origin. The
pseudocharge sector remains unchanged and the pseudospin sector of the
problem can be equivalently described by the following Hamiltonian 
\begin{eqnarray}
H^{\prime } &=&\int_{-\infty }^{\infty }dx\,\left( {2\pi }%
v_{F}O_{3}^{2}(x)-g_{2}O_{3}(-x)O_{3}(x)\right.  \notag \\
&&\left. +v_{F}B_{j}O_{j}(x)\delta (x)\right) ,  \label{fictmagnfield}
\end{eqnarray}%
where the operators $O_{j}(x)\equiv U^{\dagger }J_{j}(x)U$, with $j=1,2,3$
obey Kac-Moody relations (\ref{KacMoody}). The two Hamiltonians, (\ref{Ham_0}%
), (\ref{Ham_1}) and (\ref{fictmagnfield}), are connected by a canonical
transformation $H^{\prime }=U^{\dagger }HU$ , where 
\begin{equation*}
U=\exp i\int_{0}^{\infty }dxB_{j}J_{j}(x).
\end{equation*}%
We derive the relation between $O_{j}(x)$ and $J_{j}(x)$ in Appendix \ref%
{sec:AppA}.

Equation (\ref{fictmagnfield}) resembles the Hamiltonian for the Kondo
problem in the current algebra representation. \cite{Affleck1991} However
there are important differences. The first one is the non-local interaction $%
g_{2}$ between the incoming and outgoing waves in our case. The second
difference is the classical nature of the field $\mathbf{B}$, as opposed to
the quantum nature of the Kondo spin $\mathbf{S}$.

The matrix Green's function for chiral fermions is diagonal in pseudospin
space when defined in terms of the asymptotic eigenstates. However in terms
of the usual right- and left-movers its matrix structure is evidently given
by Eq.\ (\ref{S-parametrization}), which shows the action of the magnetic
field $S=\exp i\bm{s B}$, with $\bm{s}=\frac{1}{2}\bm{\sigma}$. We will use
this fact in Sec.\ \ref{sec:counterterms} below.

\section{RG equation for $S$-matrix}

\label{sec:RG-S}

Let us show how the renormalization group equation for the many-body $S-$%
matrix is simply reproduced in the current operator formalism. Following
Affleck and Ludwig, \cite{Affleck1991} we write the $g_{2}$ term of the $S-$%
matrix in the interaction representation as 
\begin{equation*}
T_{t}\exp 2ig_{2}\int dt\int_{a}^{L}dxJ_{3}(-x)(\widehat{R}J)_{3}(x)
\end{equation*}%
Expanding the latter object in first power of $g_{2}$, we find at given $t$ 
\begin{equation*}
-2g_{2}\int_{a}^{L}dxJ_{3}(-x)R_{3j}J_{j}(x)
\end{equation*}%
Applying OPE rules (\ref{OPE}) to the last expression one finds an operator
term 
\begin{equation*}
J_{3}(-x)J_{j}(x)\sim \frac{\varepsilon _{3jk}}{2\pi }\frac{J_{k}(0)}{-2x}
\end{equation*}%
After eliminating the distances close to the scatterer, i.e. integrating
over $x$ from $a$ to a certain $a^{\prime }$, which serves now as new lower
cutoff, we obtain the term which can be re-exponentiated to the action
giving a contribution of the form $\delta H=v_{F}\,\delta B_{k}\,J_{k}(0)$,
with 
\begin{equation*}
\delta B_{k}=\varepsilon _{3jk}R_{3j}\frac{g_{2}}{2\pi v_{F}}%
\int_{a}^{a^{\prime }}\frac{dx}{x}.
\end{equation*}%
We notice further that $\varepsilon _{3jk}R_{3j}\equiv \widetilde{B}%
_{k}=\sin 2\theta (\cos \phi ,\sin \phi ,0)$, i.e., the additional magnetic
field at the origin is parallel to the above $\mathbf{B}$, and we may
restrict the discussion to the change in modulus, $B\rightarrow B+\delta B$.
Denoting $d\Lambda =\int_{a}^{a^{\prime }}{dx}/{x}$, we obtain ${d(2\theta )}%
/{d\Lambda }=g\sin 2\theta $, or 
\begin{equation}
\frac{d\,\sin ^{2}\theta }{d\Lambda }=2g\sin ^{2}\theta \cos ^{2}\theta
\label{RG1st}
\end{equation}%
which is equivalent to Eq.\ (\ref{YMG.RGeq}).

After this verification of the previous result in the first order in $g$, we
are tempted to obtain the next-order corrections to the RG equation (\ref%
{RG1st}). It turns out however that working entirely in the current operator
formalism gives no advantage in such analysis, as we explain in the next
subsection.

\subsection{Higher orders in current operator formalism}

Let us return to the last term in (\ref{OPE}) 
\begin{equation*}
J_{j}(x)J_{k}(y)\sim \frac{\varepsilon _{jkl}J_{l}(y)}{2\pi (x-y)}.
\end{equation*}%
Using the definition (\ref{def:currents}), and free fermionic Green's
functions, one obtains e.g.\ the precise relation 
\begin{equation}
J_{2}(x)J_{3}(y)=\frac{\psi _{1}^{\dagger }(x)\psi _{2}(-y)+\psi
_{1}^{\dagger }(y)\psi _{2}(-x)+h.c.}{8\pi (x-y)},  \label{preciseOPE}
\end{equation}%
modulo the normal ordered part. Evidently, the numerator in (\ref{preciseOPE}%
) is reduced to $4J_{1}(y)$ only in the limit $x=y$. In other words, the
relation (\ref{OPE}) is the "short-distance" expansion and must be used with
caution. The meaning of the relation (\ref{OPE}) becomes clearer when we
compare the results obtained within the purely fermionic (and thus
unambiguous) approach with the results provided by the use of (\ref{OPE}).
When available, this comparison yields identical results in the limit of
large (external) times and distances. Hence, the relations (\ref{OPE})
provide a good mnemonic rule for a quick estimate of a given contribution to
a desired quantity. However, if we want to obtain concrete prefactors and
verify the renormalizability of the theory, it is better to use the standard
fermionic approach, treating the currents $J_{l}$ as pseudospin vertices.
Proceeding this way below, we use a symmetry of the problem, namely the
unitarity of the $S$-matrix related to the conservation of charge, which
leads to a pure rotation of the vector current operator after the scattering
event.

A different attempt to explore the current structure of the theory would be
the use of the equation of motion method, which we describe in the Appendices 
\ref{sec:AppC} and \ref{sec:AppC2}. If successful, this strategy would
account for the influence of interaction in all orders. 
The main difficulty arising in this approach is the nonlocal character of interaction 
(\ref{Ham_1}) and the appearance of an infinite series of coupled equations. We 
show  in Appendix \ref{sec:AppC2}, that 
the attempt to truncate this series by considering the
"most relevant" contributions leads, perhaps not surprisingly, to the
restoration of the previous result (\ref{soluRG1st}).
In order to go beyond this level of consideration,
we use the regular perturbation theory in chiral fermions in the remainder
of the paper.

\section{Corrections to conductance: perturbation theory}

\label{sec:CorrCond}

\subsection{Diagrammatic technique}
\label{sec:CorrCondA}

Let us first formulate the rules for the calculation of corrections to the
conductance in perturbation theory in the interaction $g$.

{%\color{blue}
The contributions to $G(\Omega _{m}) $ in n-th order of $g_{2}$ may be
calculated with the help of Feynman diagrams in the position-energy
representation ($\Omega _{m}$ is the external Matsubara frequency). We draw $%
n$ vertical wavy lines in parallel, each carrying the factor $-2g_{2}$, the
upper endpoint of the $i$-th line at $-x_{i}$ with pseudospin matrix $%
\frac12 \tau _{\alpha \beta }^{3}$ , the lower one at $x_{i}$ with matrix $%
\frac12 R_{3\mu }{\tau }_{\alpha \beta }^{\mu }$ attached ; $\alpha, (\beta
) $ are pseudospin indices of incoming (outgoing) fermion lines. The
external vertices are at $-y$ with matrix $\frac12 \tau _{3}$ and at $x$
with matrix $\frac12 (R\vec{\tau})_{3}$. The vertex points are connected by
Green's functions 
\begin{equation}
\mathcal{G}(x\;;\;\omega _{n})= -{i}v_{F}^{-1} \mbox{sign}
(\omega_{n})\Theta (\omega _{n}x) e^{-\omega _{n}x/v_{F}} \,,
\label{def:Green}
\end{equation}
where the $\omega _{n}$ are Matsubara fequencies $\omega _{n}=(2n+1)\pi T$.
All internal $x$ -variables are integrated on the positive semi axis. The
trace over the product of all isospin matrices in each fermion loop is taken
and a factor of $1/n!$ is applied to each n-th order diagram. The limit $%
\Omega _{m}\rightarrow +0$ is taken at the end. }

After this \ brief overview we provide further details about our
calculation. \newline
(i) From Eqs.\ (\ref{def:current}), (\ref{def:conduc}) it follows that 
% for positive external $\omega>0$ 
only two out of four correlation functions in $G(x,\omega )$ are non zero.
Namely, the chiral character of the currents (\ref{OPE}) leads to $\langle
J_{3}(x,t)J_{3}(y,0)\rangle _{\omega }=0$ at $x<y$. Then assuming $y>x=L$ in
(\ref{def:conduc}) we get 
\begin{equation*}
G(x,t)\sim \int dy\,\langle \lbrack J_{3}(-x,t)+\widetilde{J}%
_{3}(x,t),J_{3}(-y,0)]\rangle
\end{equation*}%
The first term here gives a contribution $G^{(1)}(\omega \rightarrow 0)=1/2$%
, it is not affected by interaction. The second term is $G^{(2)}(\omega
\rightarrow 0)=R_{33}/2=\cos (2\theta )/2$ for free electrons, so that $%
G=G^{(1)}+G^{(2)}=\cos ^{2}\theta =|t|^{2}$ as expected.

(ii) It is this second correlation function, $\sim \langle \widetilde
J_3(x,t) J_3(-y,0)\rangle$, which undergoes renormalization. Let us denote
this quantity by $Y$ : 
\begin{equation}
Y \equiv 2 G -1= R_{33} = \cos 2\theta  \label{def:Y}
\end{equation}
The decrease of the conductance can be viewed as the increase of the
rotation angle $\theta$. Below we discuss the corrections to $Y$.

(iii) In order $g^{n}$, we draw a diagram which contains $2n+2$ current
vertices. Two currents are external, and $2n$ come from $n$ pairs of
currents in each interaction term. We fix the coordinates of the interaction
terms and calculate the diagrams. In the limit of external frequency $\Omega
\rightarrow 0$ each diagram is $\propto \Omega ^{m}$ with $m\geq 1$, the
integration over external $y$ in (\ref{def:conduc}) reduces this power $%
m\rightarrow m-1$ so that only the linear-in-$\Omega $ contribution survives
in this limit ; see also paragraph (viii) below. After this procedure we
have \textit{finite contributions}, dependent on the positions of
interaction points, $x_{i}$, and should integrate over these positions.

(iv) These last integrals over $x_i$ ($i=1,\ldots n$), if taken over the
entire semiaxis $(0,\infty)$, \textit{diverge logarithmically} and it is
precisely at this step when we have to introduce the cutoff, as we discuss
at length below.

(v) Each current vertex has matrix structure and is $\sigma _{3}$ for a
current at negative coordinate (before scattering) and $\cos 2\theta \sigma
_{3}+\sin 2\theta \sigma _{2}$ for positive coordinate (after scattering).
We denote these matrices by $\hat{V}_{j}$. Corrections are generated by
connecting all current vertices by Green's functions. Technically, we employ 
\textit{Mathematica} and generate all possible permutations $%
\{j_{1},j_{2},\ldots ,j_{2n+2}\}$ of vertices and make one fermionic loop,
connecting these vertices in given order. The advantage of current
representation is that the information about the scattering is encoded in
Pauli matrices and the kinetics is given by chiral Green's functions (\ref%
{def:Green}). The matrix structure produces an overall prefactor before the
diagram of the form 
\begin{equation*}
\frac{1}{2}Tr[\hat{V}_{j_{1}}.\hat{V}_{j_{2}}.\cdots .\hat{V}_{j_{2n+2}}]
\end{equation*}

(vi) Other corrections are generated when $2n+2$ current vertices are
connected by more than one fermionic loop. We denote these structurally
different sets of diagrams by listing the number of current vertices in
internal loops. In the zeroth order of $g$ we have the trivial set $\{2\}$,
in the first order of $g$ we have only a set $\{4\}$. In second order of $g$
we have two sets, $\{6\}$ and $\{4,2\}$. In the third order of $g$ we have
four different sets : $\{8\}$, $\{6,2\}$, $\{4,4\}$ and $\{4,2,2\}$. Some of
these diagrams are shown in the Figures \ref{fig:diag1order} and \ref%
{fig:3rd}.

(vii) To exclude double counting of different graphs in this procedure, we
should fix one element in each loop. We always fix one external vertex as a
first element in the first loop. In addition, we assume that the coordinates
of internal vertices are ordered, i.e.\ $L>x_1>x_2>x_3>0$, etc. Then. e.g.\
the number of graphs in the $\{8\}$ set becomes $7!=5040$. In the $\{6,2\}$
set, one has $7!/2$ graphs, with factor $2$ stemming from the necessity to
fix one element in the second loop. Similarly, the set $\{4,4\}$ is obtained
by taking $7!$ permutations, then partitioning each permutation into two
parts of length 4, summing all possible diagrammatic contributions and
dividing the result by factor $4$ to account for the necessity to fix one
element in the second loop.

(viii) We do not explicitly require that both external vertices belong to
the same loop, e.g. in the diagrams of the $\{4,2\}$ type. However, before
the final integration over the internal coordinates, we seek contributions,
which are proportional to the first power of external frequency $\Omega_n$,
this linear dependence coming from two external vertices belonging to the
same loop. If these vertices belong to two different loops, then such
diagram is proportional at least to $\Omega_n^2$. After analytic
continuation $i\Omega_n \to \Omega$ and (non-logarithmic) integration over
interaction region $(-L,L)$ we have an extra factor, $\Omega L$, which is
small in the limit $\Omega \to 0$. In other words, we neglect the screening
of external field, which is given by RPA series where fermion polarization
loops are (1-reducibly) connected by interaction lines. It is allowed when
measurement times, $\sim \omega^{-1}$, are large compared to times of travel
through the interacting region, $L/v_F$. Thus we find the value unity for
the conductance in the absence of the barrier. \cite%
{Maslov1995,Safi1995,Oreg1995}

%%%%%%%%%%%%%%%%%%%%%%%%%%%%%%%%%%

\subsection{From perturbative corrections to the RG equation}

In this subsection we discuss the general form of corrections to the
conductance and outline basic ideas of the renormalization group approach,
which we will use below.

Starting with the conductance $0<G<1$ in the non-interacting limit, the
corrections $\delta G$ induced by the interaction take the form of a power
series in $g$. More precisely, we discuss the equivalent quantity $Y=2G-1$ ,
equal to $\cos 2\theta $ in the limit $g\rightarrow 0$ . We denote the
quantity $Y$ , renormalized by interaction effects, by $Y_{r}$ and recall
that it may be represented as a power series in $g$ . The coefficients of
this series contain terms varying as powers of $\Lambda =\ln (L/L_{0})$ (at
zero temperature) or $\Lambda =\ln (T_{0}/T)$ at finite temperature $T$ in
the limit of large $L$ . Here $L_{0}$ and $T_{0}$ are ultraviolet cutoff
parameters. Accordingly, $Y_{r}$ may be represented as a double series, 
\begin{eqnarray}
Y_{r} &=&Y_{0}+b_{11}g\Lambda  \notag \\
&&+b_{22}g^{2}\Lambda ^{2}+b_{21}g^{2}\Lambda  \notag \\
&&+b_{33}g^{3}\Lambda ^{3}+b_{32}g^{3}\Lambda ^{2}+b_{31}g^{3}\Lambda +\ldots
\label{expansionLog}
\end{eqnarray}%
with $b_{ij}$  functions of $Y$. The terms of the form $(g\Lambda )^{n}$
in (\ref{expansionLog}) are conventionally called \textquotedblleft
leading\textquotedblright\ contributions, and the terms of the form $%
g^{n}\Lambda ^{m}$ are called \textquotedblleft
subleading\textquotedblright\ for $n>m$. The first term $Y_{0}$ is the sum
of all scale independent contributions

\begin{equation}
Y_{0}\equiv Y_{r}(\Lambda =0)=Y+b_{20}g^{2}+b_{30}g^{3}+\ldots
\label{shifted}
\end{equation}

Naively speaking, the correction should be small $\delta Y=Y_{r}-Y\ll Y$, in
order to be considered as a correction. More precisely, the series (\ref%
{expansionLog}) should be converging. When all $|b_{ij}|\sim 1$ and $\Lambda
\gg 1$, this convergence is achieved for $g\Lambda \ll 1$ which in turn
imposes stricter conditions on the smallness of $g$. However, assuming that $%
Y_{r}(g,\Lambda )$ is an analytic function of $\Lambda ,$ we expect that an
analytical continuation from the region of small $\Lambda $ to large $%
\Lambda $ should be possible \cite{Suslov-review} . Then we can relax the
requirement of the smallness of $g$. Rearranging the above series we write 
\begin{equation}
Y_{r}(\Lambda )=Y_{0}+\beta _{1}\Lambda +\frac{\beta _{2}}{2}\Lambda ^{2}+%
\frac{\beta _{3}}{3!}\Lambda ^{3}+\ldots  \label{eq:rearrange}
\end{equation}
where the coefficients $\beta _{i}$ are functions of $Y$ .

If the theory is renormalizable, the following scaling relation holds: 
\begin{equation*}
Y_{r}(g,Y,T)=F(g,T/\Theta )
\end{equation*}%
where $\Theta =\Theta (g,Y)$ is the correlation temperature of the problem
(an analogous relation holds for the scaling with $L$ at zero temperature).
The scaling property implies the existence of a renormalization group
equation 
\begin{equation}
\frac{\partial Y_{r}(g,Y,\Lambda )}{\partial \Lambda }=\beta
(g,Y_{r}(\Lambda ))  \label{eq:RG}
\end{equation}%
where $\beta (g,Y_{r})$ is the so-called $\beta $-function. To obtain $\beta 
$ from perturbation theory we take the first derivative of (\ref%
{eq:rearrange}) with respect to $\Lambda $ and put $\Lambda =0$ 
\begin{equation}
\left. \frac{dY_{r}(g,\Lambda )}{d\Lambda }\right\vert _{\Lambda =0}=\beta
_{1}(g,Y)=\beta (g,Y_{0})  \label{def:BetaFun}
\end{equation}%
In the last equation we have replaced $Y$ by $Y(Y_{0})$ , obtained by
inverting the series (\ref{shifted}) for $Y_{0}$ in powers of $Y$, which
transforms the function $\beta _{1}(g,Y)$ into $\beta (g,Y_{0})$ . Now we
see that (\ref{def:BetaFun}) is nothing else but the RG equation (\ref{eq:RG}%
) at $\Lambda =0$ where $Y_{r}=Y_{0}$. Consequently, the function $\beta
(g,Y_{0})$ is the true $\beta $-function.

To check whether the scaling and therefore the RG equation holds, we may
integrate (\ref{eq:RG}) to recover the perturbation series, by employing the
inverse function 
\begin{equation}
\Lambda =\int_{Y}^{Y_{r}}\frac{dY}{\beta (g,Y)}
\end{equation}%
Alternatively, noting that $\frac{d}{d\Lambda }=\beta (Y)\frac{d}{dY}$ we
may obtain the Taylor expansion in powers of $\Lambda $ directly in the form 
\begin{eqnarray}
Y_{r}(\Lambda ) &=&Y+\Lambda \beta (Y)+\frac{1}{2}\Lambda ^{2}\beta (Y)\beta
^{\prime }(Y)  \notag \\
&+&\frac{1}{6}\Lambda ^{3}\beta (Y)[\beta (Y)\beta ^{\prime }(Y)]^{\prime
}+\ldots  \label{BetaTaylor}
\end{eqnarray}%
with a prime denoting the differentiation with respect to $Y$.

In what follows we, firstly, verify the structure (\ref{BetaTaylor}) of the
theory up to third order in $g$, using computer algebra. Thus we show the
applicability of the RG approach. Secondly, on the basis of this analysis we
propose a method of resummation of the most important contributions to the $%
\beta $-function in all orders of $g$.

But before presenting these results we return to the significance of the
scale independent terms in the perturbation theory (the difference between $%
Y_{0}$ and $Y$), as this problem has been sometimes overlooked in the
literature. In the first order of $g$ we can absorb such finite terms into
the definition of $\Lambda $. However, after this fixing of the definition
of $\Lambda $, higher order scale independent terms cannot be removed by
redefinition. For example, we should expect a term $b_{21}g^{2}(\Lambda +c)$
with $c\sim 1$ in the second line of (\ref{expansionLog}). It is clear that
in the third order we can find a corresponding multiplicative contribution $%
\sim b_{11}b_{21}g^{3}\Lambda (\Lambda +c)$, which contains a linear-log
term and thus affects the definition of the $\beta -$function in the form of
(\ref{def:BetaFun}). Evidently, such terms arise only at the level of $g^{3}$%
, or \textquotedblleft in the third loop\textquotedblright , and do not
affect the leading-log contributions. As this problem is of principal
importance, we present in the following a second line of derivation of the $%
\beta $-function.

Within the Callan-Symanzik (CS) formulation of the RG method \cite%
{Ramond1989,Higashijima1980} one may start with the perturbation series of $%
Y_{r}$ in terms of the bare conductance $Y$ , (\ref{expansionLog}). The
latter relation is inverted and $Y$ is expressed in terms of the
\textquotedblleft running\textquotedblright\ parameter $Y_{r}$ 
\begin{eqnarray}
Y &=&Y_{r}+\bar{b}_{11}g\Lambda  \notag \\
&&+\bar{b}_{22}g^{2}\Lambda ^{2}+\bar{b}_{21}g^{2}\Lambda  \notag \\
&&+\bar{b}_{33}g^{3}\Lambda ^{3}+\bar{b}_{32}g^{3}\Lambda ^{2}+\bar{b}%
_{31}g^{3}\Lambda +\ldots  \label{expansionCS}
\end{eqnarray}%
with $\bar{b}_{ij}$ functions of $Y_{r}$. The expansion (\ref{expansionCS})
is the essence of the CS formulation, appropriate for our problem. Here $Y$
is understood to be a function of the two variables $Y_{r}(\Lambda )$ and $%
\Lambda $ . Since the quantity $Y$ is independent of $\Lambda $ we have the
relation 
\begin{equation}
0=\frac{dY}{d\Lambda }=\frac{\partial Y}{\partial \Lambda }+\frac{\partial Y%
}{\partial Y_{r}}\frac{\partial Y_{r}}{\partial \Lambda }
\end{equation}%
or 
\begin{equation}
\beta =\frac{\partial Y_{r}}{\partial \Lambda }=-\frac{\partial Y/\partial
\Lambda }{\partial Y/\partial Y_{r}}  \label{betaCS}
\end{equation}%
In practice, the calculation of the $\beta $-function from (\ref{betaCS})
requires the calculation of the coefficients $\bar{b}_{ij}$ from
perturbation theory, which is analogous to the transformation of $\beta
_{1}(Y)$ to $\beta (Y_{0})$ discussed above. If everything is done
correctly, then the expression (\ref{betaCS}) does not depend on $\Lambda $
explicitly.

The CS procedure described above is different from the Gell-Mann--Low (GL)
formulation of RG. In the latter one considers the value of conductance used
in the calculation to be defined at the scale $\Lambda $ ; which means $%
Y_{r} $ in our notation. To eliminate the large terms involving $\Lambda $
one introduces counterterms into the Hamiltonian (cf.\ below), in order to
guarantee a finite value of the bare conductance, with a resulting equation
similar to (\ref{expansionCS}). Usually the insertion of the counterterms is
done in the \textquotedblleft minimal subtraction\textquotedblright\ scheme,
meaning that scale independent contributions are omitted. As we show in
Appendix \ref{sec:NU}, this GL formulation leads to ambiguities in the
definition of the $\beta $-function at the level of the third loop, in
contrast to the CS formulation or the scaling formulation presented in Eq.\ (%
\ref{def:BetaFun}).

\subsection{Summary for the zero temperature case}

In the perturbation theory calculation the logarithmic factors appear at the
last step, when integrating over the points of interaction. Prior to this
step the calculations are straightforward. We proceed first with the zero
temperature case. In this case the sums over discrete Matsubara frequencies
transform into integrals over imaginary energies, which are rather easily
done by computer symbolic computation.

\begin{figure}[tb]
\includegraphics[width=8.5cm]{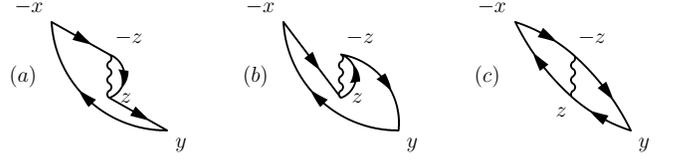}
\caption{ Three Feynman diagrams, depicting the first order in $g_{2}$
contribution to the conductance. The other three diagrams are obtained from
these by reversing the direction of fermionic propagation. }
\label{fig:diag1order}
\end{figure}

Each diagram contains a prefactor $\Omega e^{-\Omega (x+y)}$ with the two
external vertices at $x,y>0$. Integrating over $y$ and putting $\Omega
\rightarrow 0$ makes this prefactor unity. Equivalently, we may keep only
the first, linear-in-$\Omega $, term in each diagram and let $\Omega =0$ in
its remaining part. In the first order of $g$ the correction is given by the
diagrams in Fig.\ \ref{fig:diag1order}. The vertex correction part in Fig.\ %
\ref{fig:diag1order}c cancels and the two first diagrams give an expression
which contains $1/x_{1}$. This should be integrated over the point of
interaction $x_{1}$ in the limits $(a,L)$ which produces a factor 
\begin{equation}
\Lambda _{0}=\ln L/a  \label{def:Lambda0}
\end{equation}
Similarly, in the second order of $g$ we have two contributions, $%
(x_{1}x_{2})^{-1}$ and $(x_{1}+x_{2})^{-2}$, which lead to $\Lambda _{0}^{2}$
and $\Lambda _{0}$, respectively. We provide further details in Appendix \ref%
{sec:AppD} and list here only the summary of our calculation.

Grouping corrections in powers of $g$, as in (\ref{expansionLog}) we have
for the renormalized quantity $Y_r$:

\begin{eqnarray}
Y_{r} &=&Y-c_{\{4\}}{g\Lambda }(1-Y^{2})  \notag \\
&+&g^{2}Y(1-Y^{2})[-c_{\{6\}}(\Lambda ^{2}-a_{4})+c_{\{4,2\}}(\Lambda
-a_{1})/2]  \notag \\
&+&g^{3}(1-Y^{2})\Big[-c_{\{8\}}(\Lambda ^{3}-3\Lambda a_{4})(Y^{2}-1/3) 
\notag \\
&+&c_{\{8\}}(1-Y^{2})\Lambda a_{2}+c_{\{6,2\}}Y^{2}\Lambda (\Lambda -a_{1}) 
\notag \\
&-&c_{\{4,4\}}(1-Y^{2})(\Lambda ^{2}/4-a_{3}\Lambda )  \notag \\
&-&c_{\{4,2,2\}}(1+Y^{2})\Lambda /4\Big]  \label{eq:corY}
\end{eqnarray}%
where $\vec{a}=(2\ln 2,\pi ^{2}/12,(\ln 2-1)/2,0)$. All $c_{\{\cdot \}}=1$
and these coefficients are shown only for reference to the underlying group
of diagrams, according to the above conventions.

Using the above CS scheme we come after some algebra to the RG equation ${%
dY_{r}}/{d\Lambda }=\beta (Y_{r})$ with 
\begin{eqnarray}
\beta (Y) &=&(1-Y^{2})\left[ -g+g^{2}\frac{Y}{2}-g^{3}\frac{Y^{2}+1}{4}%
\right]  \notag \\
&+&c_{3}g^{3}(1-Y^{2})^{2}+O(g^{4})  \label{eq:beta-general} \\
c_{3} &=&a_{2}+a_{3}-a_{4}=\frac{\pi ^{2}}{12}-\frac{1-\ln 2}{2}  \notag
\end{eqnarray}%
Here the value of $c_{3}$ is that of the $T=0$ case (cf.\ below).
This value differs from our previously reported result 
\cite{Aristov2008}, which accounted only for the set $\{8\}$ of third 
order diagrams. Here we found an aditional contribution from the $\{4,4\}$ set. 

It is important to keep the scale independent contributions in the terms $%
\{6\}$ and $\{4,2\}$ in (\ref{eq:corY}) in order to have the agreement with
the result by Kane and Fisher in the limits of weak scattering and weak
tunneling $Y^{2}\rightarrow 1$.

One can check that the terms with higher powers of $\Lambda $ in (\ref%
{eq:corY}) are reproduced by Eqs.\ (\ref{BetaTaylor}) and (\ref%
{eq:beta-general}). In fact, this applicability of the RG equation is
manifested already in the absence of $\Lambda $ in the right-hand side of (%
\ref{eq:beta-general}). This means that, when discussing diagrammatic
contributions in $n$th order in $g$ we should concentrate only on the
least-leading logarithm, i.e. the terms $\sim g^{n}\Lambda $. These
linear-in-$\Lambda $ contributions arise in two qualitatively different
cases.

\textit{In the first case}, the entire Feynman diagram is proportional to a
single logarithm. This happens with the three first terms in (\ref%
{eq:beta-general}) stemming from diagrams with the \textit{maximum number}
of fermionic loops. In the above notation, these are the sets $\{4\}$, $%
\{4,2\}$, $\{4,2,2\}$ for corrections in order $g$, $g^{2}$, $g^{3}$,
respectively. We discuss these corrections in the next section. The finite
terms, $O(1)$, do not contribute to the $\beta $-function as can be checked
by the absence of coefficient $a_{1}$ in (\ref{eq:beta-general}).

\begin{figure}[t]
\includegraphics[width=8.5cm]{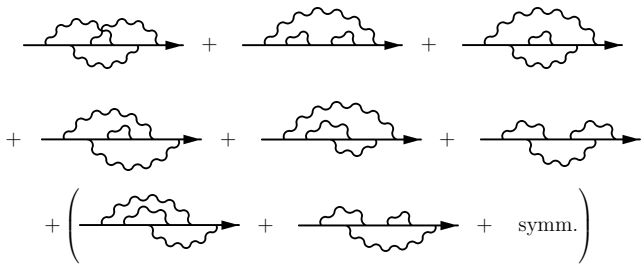}
\caption{Skeleton Feynman graphs, leading to lowest order logarithmic
contribution, $g^3 \Lambda$, beyond ladder series. }
\label{fig:3rd}
\end{figure}

In addition to these sets, there are contributions in the third order, whose
leading divergence is also linear logarithmic. They are depicted by a first
skeleton graph in Fig.\ \ref{fig:3rd}, with the \textit{maximally crossed}
wavy lines of interaction. These graphs provide for one half of the value of 
$a_2$ in Eqs.\ (\ref{eq:corY}), (\ref{eq:beta-general}).

\textit{In the second case}, the linear-log contributions arise as
accompanying weaker divergence in the stronger divergent graphs. In simple
cases, it may be a combination $\Lambda ^{3}+\Lambda O(1)$ ; in more
complicated situations, it may be a stronger singularity at internal points,
e.g. $\sim \int dz_{1}/(z_{1}-z_{2})^{-1}$ which is eventually removed upon
symmetrization, whereas the subleading log-divergence survives and
contributes to (\ref{eq:corY}), cf.\ \cite{Ludwig2003}. This second type of
linear-log contributions happens in the remaining graphs in Fig.\ \ref%
{fig:3rd} and provides for the rest of the value of the last term $c_3$ in (%
\ref{eq:beta-general}).

\subsection{Finite temperatures}

Let us discuss the effects of temperature, first in the lowest order in $g$.
At $T=0$ the large logarithm of the theory (\ref{def:Lambda0}) came from the
integration $\int dx_{1}/x_{1}$ over the range of interaction $[a,L]$. When
the integration over internal frequencies is replaced by a summation over
Matsubara frequencies, we find that the integrand contains the fermionic
Green's function $\mathcal{G}(2x,t=0)$. of the form 
\begin{equation}
\mathcal{G}(2x)=\frac{T}{2v_{F}\sinh (2\pi Tx/v_{F})}\leftrightarrow \frac{1%
}{4\pi x}.
\end{equation}%
with an appropriate change in the definition 
\begin{equation}
\Lambda _{0}\rightarrow 4\pi \int_{a}^{L}dx\;\mathcal{G}(2x)=\ln \frac{\tanh
L/\xi }{\tanh a/\xi }\equiv \Lambda
\end{equation}%
with the temperature correlation length, $\xi =v_{F}/\pi T$. Explicitly we
have 
\begin{equation}
\Lambda = \ln \coth (a/\xi) \simeq \ln \frac{\xi}{a} \simeq \ln \frac {T_0}{T%
},  \label{def:Lambda}
\end{equation}
where the last relation takes place at larger temperatures, when $L \gg \xi$%
. Here we introduced the ultra-violet energy cutoff $T_0 = v_F/\pi a$; one
may think of $a \sim k_F^{-1}$ so that $T_0 \sim E_F$. The results presented
below refer to this case of relatively high temperatures, $T_0\gg T\gg v_F/L 
$.

We calculated the terms of perturbation theory by means of computer algebra,
similarly to the $T=0$ case above. The summary of the result is given by the
same expression (\ref{eq:corY}), but the subleading coefficients are now
given by $a = (3\ln 2 ,\pi^2/12, \ln 2 -1/2, \pi^2/24 )$. In its turn, it
leads to a different numerical value of the coefficient (\ref%
{eq:beta-general}), we obtain for $T\gg v_F/L $ : 
\begin{equation}  \label{eq:c3finite}
c_3 = \frac{\pi^2}{24} + \ln 2-\frac{1}{2} \;.
\end{equation}
Let us comment on this result.

We observe that the diagrams with the maximum number of fermionic loops lead
to expressions linear in $\Lambda $ with the same prefactors as in the $T=0$
case. They generate the same part of the $\beta$ function, given by the
first line in (\ref{eq:beta-general}), and correspond to one-loop RG
approximation. In the next section we show that these terms form a series
which can be resummed in all orders of $g$.

The maximally crossed diagrams of the third order which had only linear-log
divergence at $T=0$ lead to the same $a_2/2$ coefficient in (\ref{eq:corY})
at $T\gg v_{F}/L$.

Let us now consider the modifications in the second order. The complete
expression for the diagrammatic set $\{6\}$ is given by the expression (\ref%
{TempCorr(6)}) which shows that $a_4$ in Eq.\ (\ref{eq:corY}) is a smooth
finite function of temperature.

At finite $T$ each particular diagram in order $g^{3}$ can lead to a rather
complex expression, and the resulting sum of all diagrams becomes too
complicated for a brute-force approach. The analysis can be best done in the
following way. We add all the terms of a given structural set of diagrams,
thus removing the internal singularities at $z_{1}=z_{2}$, etc. As discussed
above the leading logarithmic behavior of the set (higher power of $\Lambda$%
) is determined by the RG equation. We may hence subtract a simpler
expression, with the same leading behavior, from the complicated general
expression.

%%%%%%%%%%%%%%%%%%%%%%%%%%%%%%%
%%%%%%%%%%%%%%%%%%%%%%%%%%%%%%%%

For instance, for the $\{8\}$ set we pick all diagrams proportional to $\cos
8\theta $, and notice that the zero-$T$ expression has a leading
contribution $(z_{1}z_{2}z_{3})^{-1}$ with a certain coefficient. Then we
expect that the finite-$T$ expression should have a leading term $8\xi
^{-3}(\sinh (2z_{1}/\xi )\sinh (2z_{2}/\xi )\sinh (2z_{3}/\xi ))^{-1}$ with
the same coefficient. We subtract this term from our expression and analyze
the degree of singularity of the remaining expression. It turns out that in
the third order after the subtraction of such leading $\Lambda ^{3}$ terms
we are left with linear-log expressions. The coefficients before these
expressions are also smooth functions of $T$, as we explain in the Appendix %
\ref{sec:AppE}.

Concluding this section, we verified the validity of the RG-equation by
checking that the higher-than-linear powers of $\Lambda $ are generated by
the linear-log terms. This holds both at $T=0$ and at $T\gg v_{F}/L$; the
resulting $\beta $-function is somewhat different in these cases and we
discuss this difference below. There is a part of the $\beta $-function,
which is identical in both cases, and stems particularly from a sequence of
diagrams with the maximum number of polarization loops. We analyze this
sequence in the next Section and show that it is possible to resum it in all
orders of $g$.

It is important to note that the last term in (\ref{eq:beta-general}) is
vanishing in the limit when the barrier becomes fully transparent or fully
reflecting, i.e. at $Y\rightarrow \pm 1$. In order to see that, let us
consider the case $(Y+1)/2=G\rightarrow 0$, then the first three terms in (%
\ref{eq:beta-general}) read $\beta \simeq 2G\left[ -g-g^{2}/2-g^{3}/2\right] 
$ and lead to power-law scaling of $G$ in accordance to resutls by Kane and
Fisher (see below). At the same time the last term in (\ref{eq:beta-general}%
) is $\propto g^{3}G^{2}$ and does not lead to a noticeable effect in the
limit considered.

It is also worth noting here, that the calculation both at $T=0$ and at $%
T\gg v_{F}/L$ confirms the relation between the subleading coefficients in
sets $\{8\}$ and $\{6,2\}$ on one hand and $\{6\}$ and $\{4,2\}$ on the
other hand. This is seen in Eq.\ (\ref{eq:corY}) for the $a_{4}$ and $a_{1}$
coefficients. Without this relation the results by Kane and Fisher for weak
and strong barrier would not be recovered.

%%%%%%%%%%%%%%%%%%%%%%%%%%%%%%%%%%%%%%%%%

\section{Counterterms in the Hamiltonian}

\label{sec:counterterms}

In the previous section we saw that the corrections to the conductance are
obtained as self-energy insertion into the fermionic loop, whereas the
vertex corrections vanish in the dc limit $\Omega \ll v_{F}/L$. This means,
in particular, that the effect of renormalization in this limit can be found
at the level of the one-particle Green's function, or by addition of
counterterms to the effective low-energy Hamiltonian.

We calculate the corrections to the matrix Green's function for the chiral
fermions in the pseudospin sector in perturbation theory. The rules of the
diagrammatic technique remain the same. This time we do not consider a
closed fermionic loop, which described the conductance above. Instead we
study the matrix Green's function $\widehat G(x,-y,\Omega _{n})$ connecting
distant points on different sides of the barrier in the chiral
representation. The initial Green's function is diagonal in pseudospin
space, and the corrections to it make it non-diagonal. We show that this
non-diagonal form corresponds to additional rotation of pseudospin near the
origin, in accordance with consideration of Sec.\ \ref{sec:RG-S}.

Employing the CS scheme we start with the bare barrier, characterized by the
parameter $\theta$ and find the renormalized value $\theta_r(\theta,
\Lambda) $ as described below. Then we determine the inverse relation $%
\theta(\theta_r, \Lambda)$ and require the independence of initial $\theta$
at the scale of consideration $\Lambda$. This provides us with the RG
equation for $\theta_r$

The prefactor in $G(x,-y,\Omega _{n})=\widehat{\mathcal{G}}\exp -\Omega
_{n}(x+y)$ is found as

\begin{eqnarray}
\widehat{\mathcal{G}} &=& 1+c_{\{4\}}\frac{g\Lambda }{2}i\sigma _{1}\sin
2\theta  \label{GreenLambda0} \\
&&-c_{\{6\}}\frac{g^{2}}{8}(\sin ^{2}2\theta \Lambda ^{2}-i\sigma _{1}\sin
4\theta (\Lambda ^{2}- b_{1}))  \notag \\
&&-c_{\{4,2\}}\frac{g^{2}}{8}i\sigma _{1}\sin 4\theta (\Lambda - b_3)  \notag
\\
&&-c_{\{8\}}\frac{g^{3}}{48}\Big(3(\Lambda ^{3}-b_{1}\Lambda )\sin 2\theta
\sin 4\theta  \notag \\
&&-i\sigma _{1}\frac{\Lambda ^{3}}{2}\sin 2\theta (9\cos 4\theta -1)  \notag
\\
&& +i 12 \sigma _{1}\Lambda \sin 2\theta ( b_{2}\sin ^{2}2\theta + b_{1})%
\Big)  \notag \\
&&+c_{\{6,2\}}\frac{g^{3}}{16}\Lambda (\Lambda -b_{3})\sin 4\theta (\sin
2\theta -2i\sigma _{1}\cos 2\theta )  \notag \\
&&+c_{\{4,4\}}\frac{g^{3}}{8}i\sigma _{1}\sin ^{3}2\theta \Lambda (\Lambda
-b_{4})  \notag \\
&&+c_{\{4,2,2\}}\frac{g^{3}}{32}i\sigma _{1}\Lambda (5\sin 2\theta +\sin
6\theta )  \notag
\end{eqnarray}%
where again all $c_{\{\cdot \}}=1$. The coefficients $b_{j},j=1,...4$ depend
on the case in consideration. For the zero temperature we find $b= (0,
\pi^2/6, 2\ln 2, 2-2\ln 2)$ and for the finite temperature $b= ( \pi^2/12,
0, 3\ln 2, 2-4\ln 2)$.

In order to establish the connection to the Hamiltonian we observe that the
above form (\ref{GreenLambda0}) can be represented as $\widehat{\mathcal{G}}
=\exp i\sigma _{1}( \theta_r- \theta) $ with

\begin{eqnarray}
\theta_r &=&\theta +\frac{g\Lambda }{2}\sin 2\theta +\frac{g^{2}}{8}\sin
4\theta (\Lambda ^{2}-\Lambda -b_{1}+b_3 )  \notag \\
&&+\frac{g^{3}}{16}\sin 2\theta \Big ( \frac43\Lambda ^{3} \cos 4\theta +
\Lambda^2(1+3\cos 4\theta)  \notag \\
&& +\Lambda ((3-4b_1 - 2b_2+2b_3-b_4)  \notag \\
&& +(1+2b_2+2b_3+b_4)\cos 4\theta) \Big) +O(g^{3} \Lambda^0)
\label{eq:thetar}
\end{eqnarray}
This means that the effect of the interaction is indeed reduced to a change
in the magnetic field $2\theta$, in accordance with Sec. \ref{sec:magfield}.

We have for the value at $\Lambda=0$ 
\begin{equation*}
\theta_0 = \theta + \frac{g^{2}}{8}(b_3-b_1)\sin 4\theta + \ldots ,
\end{equation*}
Expressing $\theta$ by $\theta_0$ in (\ref{eq:thetar}) and inverting the
series we obtain 
\begin{eqnarray}
\theta_0 &=&\theta_r - \frac{g\Lambda }{2}\sin 2\theta_r +\frac{g^{2}}{8}%
\sin 4\theta_r (\Lambda ^{2}+\Lambda )  \notag \\
&&-\frac{g^{3}}{16}\sin 2\theta_r \Big ( \frac43 \Lambda ^{3} \cos 4\theta
_r + \Lambda^2(1+3\cos 4\theta_r)  \notag \\
&& +\Lambda ((3-2b_1 - 2b_2-b_4)  \notag \\
&& +(1+2b_1+2b_2+b_4)\cos 4\theta_r) \Big) +\ldots  \label{CTermT0}
\end{eqnarray}
The difference $\theta_0 - \theta_r$, expressed in terms of $\theta_r$, is
interpreted as the counterterms to the Hamiltonian, needed to compensate the
divergent diagrammatic contributions. In our approach we do not re-calculate
diagrams with inclusion of the appropriately chosen counterterms, which
would be problematic, given the non-Abelian character of the theory and the
absence of Wick's theorem. These counterterms merely reflect the structure
of the corrections to the magnetic field, Sec.\ \ref{sec:magfield} above.

Demanding the independence of $\theta_0$ on $\Lambda$, we obtain (here $%
\theta(\Lambda)=\theta_r$) 
\begin{eqnarray}
2\frac{\partial \theta }{\partial \Lambda } &=&g\sin 2\theta -\frac{g^{2}}{4}%
\sin 4\theta +\frac{g^{3}}{4}\sin 2\theta (1+\cos ^{2}2\theta )  \notag \\
&&-c_{3}g^{3}\sin ^{3}2\theta +O(g^{4})  \label{eq:RGtheta}
\end{eqnarray}
where 
\begin{eqnarray}
c_{3} &=& \frac 14 (2b_1+2b_2+b_4)  \notag \\
&=& \frac{\pi^2}{12} + \frac{\ln 2}{2} -\frac{1}{2}\simeq 0.67 , \quad T=0 
\notag \\
&=& \frac{\pi^2}{24} + \ln 2-\frac{1}{2}\simeq 0.60 , \quad T\gg v_F/L 
\notag
\end{eqnarray}
which agrees with the above Eqs.\ (\ref{eq:beta-general}), (\ref{eq:c3finite}%
). This coincidence confirms that the vertex corrections to the conductance
are unimportant in the considered dc limit.

Any method of rendering the logarithmically divergent integrals finite is
called a scheme of regularization. As we explained above the contributions
linear in $\Lambda$ get additional contributions in third order of $g$ from
the scale independent terms appearing in the second order of $g$. We removed
this ambiguity by using the scaling method or the appropriately devised CS
scheme. A part of the answer is independent of the regularization scheme and
technically is obtained by summing the one-loop contributions to $\beta$%
-function. One loop means here one independent integration over the internal
energy, leading to a linear logarithm. Indeed the summation of the ladder
series in Sec.\ \ref{sec:ladder} amounts to a finite renormalization of the
effective interaction constant, $g \to \widetilde g$. In addition to that,
the quantities calculated above in the order $g^3$ included linear-log
contributions stemming from diagrams with three independent energy
integrations (Matsubara frequency summations). Particularly, the set of
diagrams $\{ 4,4\}$ contains diagrams with vertex corrections to the
polarization loops in the ladder series Fig.\ \ref{fig:ladder}, which
contribute to the regularization-dependent coefficient $b_4$ in Eq.\ (\ref%
{GreenLambda0}) and $c_3$ in (\ref{eq:RGtheta}).

%%%%%%%%%%%%%%%%%%%%%%%%%%%%%%%%%%%%%%%%%

\section{Infinite resummation of terms in the $\protect\beta-$function}

\label{sec:ladder}

We observe that the first three terms in (\ref{eq:beta-general}), (\ref%
{eq:RGtheta}) are given by diagrams with maximum number of fermionic loops.
These terms are $\propto (1-Y^{2})$ and therefore are important at any value
of the conductance. It is possible to resum all the contributions of this
type, as we discuss below.

\subsection{Ladder series and Wiener-Hopf equation}

\begin{figure}[tp]
\includegraphics[width=8.5cm]{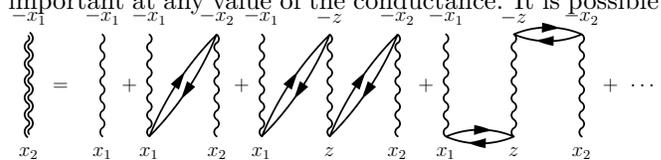}
\caption{Ladder series, $\bar{L}(x_{1},x_{2};\protect\omega _{n})$,
describing the combined effect of interaction and barrier, and leading to a
linear-in-logarithm $\ln (T_{0}/T)$ contribution to conductance. }
\label{fig:ladder}
\end{figure}

The above discussed linear-log contributions to the conductance with the
maximum number of fermionic loops, can be viewed as a \textquotedblleft
dressing\textquotedblright\ of the wavy interaction line $g$\ \ in Fig.\ \ref%
{fig:diag1order}a, Fig.\ \ref{fig:diag1order}b by fermionic polarization
loops. This is shown schematically in Fig.\ \ref{fig:ladder}, with the
result of the summation denoted by $\widetilde{g}$ and depicted by\ a double
wavy line.

We show below that the result of this summation is scale independent
(without logarithmic terms), so that using the double wavy line in
first-order contributions Fig.\ \ref{fig:diag1order}a, Fig.\ \ref%
{fig:diag1order}b will produce a single logarithmic factor, while containing
all powers of $g$.

%%%%%%%%%%%%%%%%%%%%%%%%%%%%%
%%%%%%%%%%%%%%%%%%%%%%%%%%%%%

In the usual situation, the RPA summation of Fig.\ \ref{fig:ladder} is
trivial after Fourier transform. In our case of chiral fermion
representation, we have left-right asymmetry of the vertices whose value in
pseudo-spin space depends on the coordinate. Hence we always integrate over
the semi-axis only, which leads, as shown below, to an integral equation of
the Wiener-Hopf type, instead of a simple Fourier convolution. Still, the
ladder equation for the dressed interaction can be written and solved rather
simply.

Let us denote the interaction between the points $-x$ and $y$ by $\bar
L(x,y;\omega_n)$ where $\omega_n$ is the Matsubara frequency along the line.
Let us define the vector $X(x,y;\omega_n)$ in the following way 
\begin{equation}
X(x,y;\omega_n) = 
\begin{pmatrix}
\bar L(x,y;\omega_n) \\ 
\bar L(x,-y;\omega_n)%
\end{pmatrix}%
\end{equation}

Then the ladder summation, Fig.\ \ref{fig:ladder}, is expressed for $%
\omega_n >0 $ by the integral equation :

\begin{widetext}

  \begin{eqnarray}
  X(x,y;\omega_n) &=&
  -4\pi g \delta(x-y)
  \begin{pmatrix}   1  \\    0    \end{pmatrix}
  %\nonumber \\ &-&
   - g \omega_n \int dz\,
    \begin{pmatrix}
  Y e^{-\omega_n(x+z)} & \Theta(x-z) e^{-\omega_n(x-z)}\\
  \Theta(z-x) e^{-\omega_n(z-x)} &0
  \end{pmatrix} .X(z,y;\omega_n)
 \label{equLadMatrix}
  \end{eqnarray}
where we used the fact that the polarization loop in the $(\omega_n,x)$ representation is 
$\langle J_3(x,\tau)  J_3(y,0) \rangle_{\omega_n} = (4\pi)^{-1}\omega_n  \Theta(\omega_n x) 
e^{-\omega_n x  }$. 
Here and below the integration over coordinates is over the interval
$(a,L)$, it is convergent and may be extended to $(0,\infty)$. 

Later we will need the integrated quantity $L(x;\omega
_{n})=\int d y e^{-\omega _{n}y}\;
\bar{L}(x,y;\omega _{n})$,
which obeys the integral equation
\begin{eqnarray}
L(x;\omega) &=&-ge^{-\omega x}\left[ 4\pi +\omega
(Y+\frac{g}{2})
\int d ze^{-\omega z}L(z;\omega)\right]
%\nonumber \\ &+&
+\frac{1}{2}g^{2}\omega \int d z\;\;e^{-\omega
|x-z|}L(z;\omega)  \,.
\label{equLad}
\end{eqnarray}

\subsection{Solution and its properties}
The above equation for $L(x;\omega)$ is an integral
equation of Wiener-Hopf type, and its solution can be found as follows. We
differentiate both sides of Eq.\ (\ref{equLad}) twice with respect to $x$ and find
that $d^2L(x;\omega)/dx^2 = \omega^2 (1-g^2) L(x;\omega)$.
Demanding that $L(x \to \infty;\omega)= 0$ we seek a solution in the
 form $L(x;\omega)=  C \exp(- \omega x \sqrt{1-g^2})$. Substituting
the latter form into (\ref{equLad}), we determine the constant $C$ and find 
% \[
%  C=  -4\pi  g ({1+\sqrt{1-g^2} })/({1+\sqrt{1-g^2} + g Y}) \,.
%\]
\begin{equation}
L(x;\omega)=  -4\pi  g \frac{1+\sqrt{1-g^2} }{1+\sqrt{1-g^2} + g Y} 
e^{- \omega x \sqrt{1-g^2} }\,.
\label{sol:Lx}
\end{equation}
Now we are in a position to determine the linear-log correction to
the conductance stemming from the diagrams Fig.\ \ref{fig:diag1order} 
where the wavy line of interaction is replaced by the double wavy line.

The equation reads :

\begin{eqnarray}
G^{(L)} &=&\frac{1-Y^{2}}{4}T^{2}\sum_{\epsilon ,\omega
}\int d x_{1} d x_{2} d y\; \bar{L}(x_{1},x_{2};\omega
)\mathcal{G}(x_{1}-x;\epsilon )
% \nonumber \\&\times &
\mathcal{G}(-x_{1}-x_{2};\epsilon -\omega )\mathcal{G}
(x_{2}-y;\epsilon )\mathcal{G}(y+x;\epsilon +\Omega )
\, , \label{LadContrib}
\end{eqnarray}
\end{widetext}where the superscript $L$ in $G^{(L)}$ denotes ladder
summation. Taking here the limit $\Omega \rightarrow 0$ we find 
\begin{equation}
G^{(L)}=\frac{-2g(1-Y^{2})}{1+\sqrt{1-g^{2}}+gY}\Lambda _{0}\equiv -%
\widetilde{g}(1-Y^{2})\Lambda _{0}\,.  \label{ladder:result}
\end{equation}%
Here we defined the renormalized interaction constant $\widetilde{g}$. We
observe that 
\begin{eqnarray}
\widetilde{g} &= & 1- K , \qquad Y=1 \\
&= & K^{-1} -1 , \quad Y=-1  \notag
\end{eqnarray}
with the Luttinger parameter $K=\sqrt{(1-g)/(1+g)}$.

In fact, the expression (\ref{LadContrib}) already assumes the limit $\Omega
\to 0$, because the prefactor $1-Y^{2}$ is obtained by adding two
contributions, one shown in Fig.\ \ref{fig:diag1order}b ($\propto \Omega
(1-2Y^2)$) and another in Fig.\ \ref{fig:diag1order}a with inverted
direction of fermionic lines ($\propto \Omega $).

The importance of the limit $\Omega \rightarrow 0$ in the derivation of (\ref%
{ladder:result}) is further illustrated by the solution $\bar{L}(x,y;\omega
_{n})$ of the initial Eq.\ (\ref{equLadMatrix}) rather than Eq.\ (\ref%
{equLad}). After a long and straightforward calculation, similar to the one
described above, we find 
\begin{eqnarray}
\frac{1}{4\pi g}\bar{L}(x,y;\omega _{n}) &=&-\delta (x-y)-\frac{\omega
_{n}g^{2}}{2d}e^{-\omega _{n}d|x-y|}  \label{Lxy} \\
&+&\frac{\omega _{n}g^{2}}{2d}e^{-\omega _{n}d(x+y)}\frac{Y(1+d)+g}{Y(1-d)+g}
\notag
\end{eqnarray}%
with $d=\sqrt{1-g^{2}}$. The first term in (\ref{Lxy}) is the bare
interaction, the second one is the result of RPA summation, which is
independent of the barrier transparency $Y$ and finite even in the bulk, at $%
x\simeq y\rightarrow \infty $. The last term explicitly contains the
transparency $Y$ and decays away from the barrier. Integrating over $y$ we
get the above Eq.\ (\ref{sol:Lx}), and Eq.\ (\ref{Lxy}) shows that the
\textquotedblleft dressing\textquotedblright\ of the interaction constant, $%
g\rightarrow \widetilde{g}$, in (\ref{ladder:result}) is not reduced to a
simple multiplicative factor, but depends on the coordinates. Finally we
note that $G^{(L)}=0$ at $\Lambda =0$, i.e. there are no scale independent
contributions in the ladder approximation, and $Y_{0}=Y$ in that case.

\section{RG equation and its solution}

%%%%%%%%%%%%%%%%%%%%%%%%%%%%%%%%

From the solution of the Wiener-Hopf equation for the ladder series (\ref%
{ladder:result}) we then find the $\beta $-function in ladder approximation

\begin{equation}
\beta _{L}(Y)=-\frac{2g(1-Y^{2})}{1+\sqrt{1-g^{2}}+gY}
\end{equation}%
Inverting this equation and using the definition of $K$ we obtain the
symmetric form

\begin{equation}
\frac{1}{\beta _{L}(Y)} =\frac{d\Lambda }{dY}=-\frac{1}{2} \left[\frac{1}{1-K%
}\frac{1}{1-Y}-\frac{1}{1-K^{-1}}\frac{1}{1+Y} \right]  \label{eq:ladderRG}
\end{equation}%
invariant under the simultaneous exchange of $K\leftrightarrow K^{-1}$ and $%
Y\leftrightarrow -Y$ . This differential equation may be solved analytically 
\cite{Aristov2008}. We recall that it is asymptotically exact in the limit $%
|Y|\rightarrow 1$ .

Let us now turn to the first correction beyond this series in Eq.\ (\ref%
{eq:beta-general}). Applying the correction as calculated we get the
improved RG equation

\begin{equation}
\frac{dY}{d\Lambda }=-\frac{2g(1-Y^{2})}{1+\sqrt{1-g^{2}}+gY}%
+c_{3}g^{3}(1-Y^{2})^{2}  \label{eq:RG-Y}
\end{equation}%
The correction term is qualitatively different in that it vanishes
quadratically in the limit $|Y|\rightarrow 1$ and hence is irrelevant in
this limit (considered in the early work by Kane and Fisher). The duality
property still holds, i.e. the invariance of (\ref{eq:RG-Y}) under the
simultaneous change $g\rightarrow -g$ and $Y\rightarrow -Y$ (or $\theta
\rightarrow \pi /2-\theta $ in (\ref{eq:RGtheta})). Based on these
observations, we expect that the fourth-order terms beyond the ladder series
in the RG equation will be proportional to $g^{4}Y(1-Y^{2})^{2}$. One should
keep in mind that the correction terms here have been included only to order 
$g^{3}$ . We now discuss a further improvement.

We may perform yet one more partial resummation by dressing interaction
lines as $g\rightarrow \widetilde{g}$, where

\begin{equation}
\widetilde{g}=2\left( \frac{1+K}{1-K}+Y\right) ^{-1}
\end{equation}%
The $c_{3}$ term in (\ref{eq:RG-Y}) results from non-ladder diagrams.
Therefore without double-counting we can dress the interaction constant
there and write $c_{3}\widetilde{g}^{3}(1-Y^{2})^{2}$. Then (\ref%
{eq:ladderRG}) receives a correction term of the form

\begin{eqnarray}
\frac{d\Lambda }{dY} &=&-\frac{1}{2} \left[\frac{1}{1-K}\frac{1}{1-Y}-\frac{1%
}{1-K^{-1}}\frac{1}{1+Y}\right]  \notag \\
&-&c_{3}\frac{2(1-K)}{1+K+Y(1-K)}+\ldots  \label{eq:RG-inv}
\end{eqnarray}%
The term proportional to $c_{3}$ here is of order of $g$ at small $g$ and we
omitted higher order terms.

Integrating Eq.\ (\ref{eq:RG-inv}) we obtain the following result ~:

\begin{eqnarray}
\left( {T}/{T_{0}}\right) ^{2(1-K)} &=& {\Phi(G)}/{\Phi (G_{0})} \,,
\label{bestapprox} \\
\Phi(G)&=&\frac{G^{K}}{1-G} \left( K+G(1-K)\right) ^{4c_{3}(1-K)}\,,  \notag
\end{eqnarray}

\noindent we assume the initial condition $G=G_{0}=(1+Y_{0})/2=\cos
^{2}(\theta )+c_{3}\widetilde{g}_{0}^{2}\cos (2\theta )\sin ^{2}(2\theta )$
at $T=T_{0}$, and $\widetilde{g}_{0}^{-1}=\frac{1}{2}\left( \frac{1+K}{1-K}%
+\cos (2\theta )\right) $; here $c_{3}$ is given by (\ref{eq:c3finite}) for
the case $T\gg v_{F}/L$. Evidently, the renormalization stops at $T\alt %
v_{F}/L$.

\section{Summary and discussion}

\label{sec:discussion}

In this work we have presented a fermionic formulation of the transport
theory of \ interacting electrons in a onedimensional system with potential
barrier in the linear response regime. We considered the Luttinger liquid
model with interaction $g_{2}$ (forward scattering involving right and left
movers), and employed a current algebra represention, leading to a
substantial simplification of perturbation theoretical calculations. Using
the renormalization group idea in various forms we showed that the
conductance as a function of temperature (or, at zero temperature, as a
function of system length) obeys scaling, and therefore may be obtained by
integrating a renormalization group equation. The $\beta $-function of the
latter equation has been found within a ladder approximation, which becomes
exact in the limit of small or large conductance, at any interaction
strength. Corrections to the ladder approximation up to and including third
order in the interaction, and relevant at intermediate values of the
conductance, have been calculated. These correction terms require the
solution of a fundamental problem of renormalization group theory: What is
the significance of and how does one treat scale independent contributions
appearing in perturbation theory? We show that these terms have to be taken
into account, as they guarantee the scaling of the conductance and can lead
to an important redefinition of the $\beta $-function. In that respect our
work is of general interest in the context of the renormalization group
method, beyond the particular transport problem considered here.

Eq.\ (\ref{bestapprox}) is one of the central results of our paper. It is in
agreement with the previous findings \cite{Kane1992,Kane1992a,Yue1994} in
the limiting cases $G\rightarrow 1$, $G\rightarrow 0$ (i.e. $|Y|\rightarrow
1 $ ), $K\rightarrow 1 $, except for the fact that in these previous works
the conductance tends to $K$, rather than $1$ , in the limit of vanishing
barrier strength at finite temperature. In this limit ( $|Y|\rightarrow 1$ )
our result is exact for any value of $K$ . In the intermediate regime $%
G\simeq \frac{1}{2}$ there appear corrections to our result. We have
calculated these corrections in the lowest order, which is third order in
the interaction strength. The parameter $c_{3}$ determines the strength of
these corrections. We have shown that $c_{3}$ is nonuniversal in the sense
that it depends on the physical nature of the ultraviolet cutoff, e.g. the
temperature cutoff or the length cutoff. We are convinced that the scaling
functions of the temperature dependent conductance, $G(T)$, in the limit of
infinite sample length $L$ and the length dependent conductance, $G(L)$, at $%
T=0$ are slightly different.

Let us discuss the role of the factor $c_{3}$ in (\ref{bestapprox}) which is
new as compared to previous studies \cite{Kane1992,Kane1992a,Yue1994} . It
is clear that $c_{3} $ in (\ref{eq:RGtheta}) defines the renormalization of
the conductance in the intermediate region, $G\sim 1/2$, and thus determines
the coefficient relating high-temperature and low-temperature asymptotes. We
may ask the following question: what is the coefficient $C$ in the
asymptotic expression $G\simeq C\tau ^{2(K^{-1}-1)}$ ($\tau ={T}/{T_{0}}$ )
if we fix the overall temperature scale at hight $T$ as $G\simeq 1-\tau
^{2(K-1)}$ ? From Eq.\ (\ref{bestapprox}) we have 
\begin{equation}
C=K^{4c_{3}(1-1/K)},  \label{eq:matching}
\end{equation}

At this point it is appropriate to compare our findings to the
exact results obtained within the conformal field theory (CFT) approach to
the bosonized version of the problem, Sec.\ \ref{sec:BSG}. In these earlier
works the $\beta $-function of the RG equation was not discussed. However
results equivalent to those by Fendley et al.\ \cite%
{Fendley1995,Fendley1995a} are presented in a recent study by Lukyanov and
Werner \cite{Lukyanov2007}, where particularly the $\beta $-function was
obtained in a form equivalent to our (\ref{bestapprox}). According to
Lukyanov (private communication), the coefficient $c_{3}=1/4$, which
deviates from our value $c_{3}\simeq 0.6$ and we use both values for
comparison below. Unfortunately, one rarely, if ever, discusses the scheme
of regularization used in obtaining the CFT exact solutions. We have to
conclude that the regularization used by \cite{Lukyanov2007} \ is different
from ours. The choice of cutoff scheme in our work is not based on
mathematical convenience, but rather it is following naturally from
perturbation theory. Therefore, as a theoretical result to be applied to
explain experimental data, our (\ref{bestapprox}) has advantages over the
CFT results.

We saw above that passing from the backscattering amplitude in the
Hamiltonian to the conductance (i.e.\ $S$-matrix) may require ultraviolet
regularization, i.e.\ certain prescriptions of dealing with short-distance
singular quantities. Our observation is supported by the study of the free ($%
K=1$) BSG theory, \cite{Callan1994} where the authors arrive at a
current algebra formulation similar to our treatment above. They notice that
the description of the strong coupling limit $|u_{2}|\rightarrow \infty $
corresponds to taking the quantity $|\theta |\rightarrow \pi $, (see Eq.\
(2.23) there), which is related to our $|\mathbf{B}|=2\theta $ and hence to
the conductance. They also notice that the connection between $|\theta |$
and $|u_{2}|=\theta f(\theta ^{2})=\theta +\theta ^{3}O(1)$ begins to depend
on the type of regularization chosen in the third order of $\theta $. On the
other hand, given that the CFT exact solution in the interacting case $K\neq
1$ is expressed in terms of the scaling variable $|u_{2}|(T_{0}/T)^{1-K}$,
we see that the implied equation $|u_{2}|(T_{0}/T)^{1-K}=\theta f(\theta
^{2})$ corresponds in form exactly to our (\ref{bestapprox}). It would be
thus desirable to understand the particular regularization used in the CFT
solution in more detail.

%%%%%%%%%%%%%%%%%%%%%%%%%%%%%%%%%%%

\begin{figure}[t]
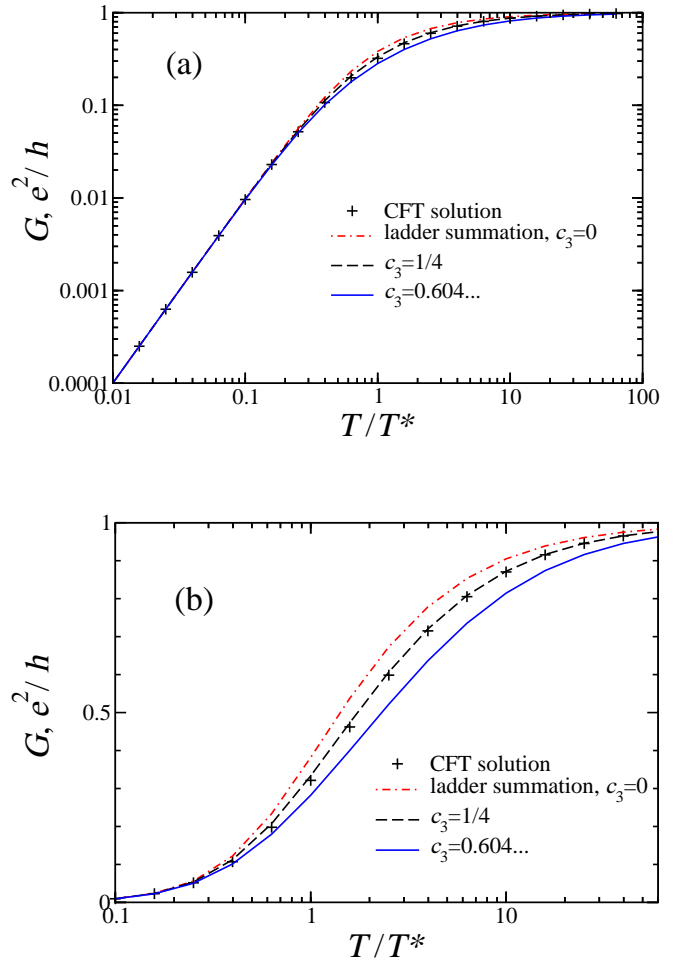

%(b)\includegraphics[width=0.9\columnwidth]{figure5b}
%\vspace{.8cm} \\
\includegraphics[width=\columnwidth]{figure5a}
\vspace{.6cm} \\
\includegraphics[width=\columnwidth]{figure5b}
\caption{(Color online) (a) Conductance as a function of temperature for $K=1/2$. The results
of ladder summation, (\protect\ref{bestapprox}) with $c_3=0$, are shown by
dashed-dotted line ; the result of Eq.\ (\protect\ref{bestapprox}), with $%
c_3 $ given by (\protect\ref{eq:RGtheta}) is shown by solid line. The result
available from CFT (multiplied by factor 2) is shown by pluses. \protect\cite%
{Kane1992,Kane1992a,Weiss1995}, together with our fit by (\protect\ref%
{bestapprox}) with $c_3=1/4$ shown as dashed line. (b) The same plot in log-linear scale.}
\label{fig:conduc}
\end{figure}
%%%%%%%%%%%%%%%%%%%%%%%%%%%%%%%%%%%

Let us now illustrate the details of the behavior of $G(T)$ in the example
of the well-known case $K=1/2$. Both high-temperature and low-temperature
asymptotes of $G$ are interesting from the theoretical viewpoint, but one
can expect the experimental variation of the conductance to be most visible
at lower $T$, when $G\alt1/2$. It is therefore useful to consider the
variation of the conductance for fixed $K$ and different values of $c_{3}$.
We take $K=1/2$ and and fix the low $T$ behavior in the form $G\simeq \tau
^{2}$, with $\tau =T/T_{\ast }$ and $T_{\ast }\sim T_{0}|r|^{2}$. We then
plot $G(\tau )$, obtained from (\ref{bestapprox}) for several values of $%
c_{3}$ in Fig.\ \ref{fig:conduc}, together with the exact solution, known
from \cite{Kane1992,Kane1992a,Weiss1995} 
\begin{equation}
G_{1/2}=1-\frac{1}{2\sqrt{3}\tau }\psi ^{\prime }\left( \frac{1}{2}+\frac{1}{%
2\sqrt{3}\tau }\right)  \label{G@K=1/2}
\end{equation}%
where $\psi ^{\prime }(x)$ the derivative of digamma function. For $K=1/2$,
and the implied CFT value $c_{3}=1/4$ we can invert (\ref{bestapprox}) and
write 
\begin{equation}
G=\frac{2\tau ^{2}}{1+2\tau ^{2}+\sqrt{1+8\tau ^{2}}}
\end{equation}

The Fig.\ \ref{fig:conduc} shows that the CFT solution lies between the
curve corresponding to pure ladder summation, $c_{3}=0$, and the one with
the above $c_{3}\simeq 0.60$. The proposed value $c_{3}=1/4$ fits the CFT
solution rather precisely. This tells us that the approximation leading to (%
\ref{bestapprox}) works rather well at the interaction strength $K=1/2$ ($%
g=3/5$) , provided the coefficient $c_{3}$ appropriate for the
regularization scheme used in the CFT solution is taken. In fact, we show in
the Appendix \ref{sec:CFT}, that the overall agreement criterion using the
coefficient $C$, Eq.\ (\ref{eq:matching}), is about 1 \% in this case. For a
further case reported ($K=1/3$ , $g=4/5$ ) \cite{Fendley1995,Fendley1995a}
the agreement is worse, about 10 \%, which might be expected in this more
strongly interacting situation.

Finally, we emphasize again that in any comparison with experiment the
correct regularization scheme is determined by the details of the
corresponding experimental setup. Indications of the possible inadequacy of
the regularization scheme used in the CFT solutions exist. For instance, we
tend to interpret the systematic deviation of the experimental data on the
conductance between the quantum Hall edge states from the CFT prediction at
low $T$ in \cite{Fendley1995a} not as a shortcoming of the experiment, but
rather as a consequence of choosing an inappropriate regularization. The
correct regularization might lead to a different value of $c_{3}$ in (\ref%
{bestapprox}), providing better agreement with experiment.

%%%%%%%%%%%%%%%%%%%%%%%%%%%%%%%%%%%

\begin{acknowledgements}
 We are grateful to I.V.\ Gornyi, D.G.\ Polyakov, P.M.\  Ostrovsky, K.A.\
Matveev, D.A.\ Bagrets, O.M.\ Yevtushenko, M.N.\ Kiselev, A.M.\
Finkel'stein, A.A.\ Nersesyan, L.I.\ Glazman and N. Andrei for various useful
discussions.
\end{acknowledgements}

\appendix 

\section{From the Hamiltonian to the S-matrix}

\label{sec:appSmatrix}

The continuum Hamiltonian (\ref{impHam}) corresponds to a lattice model 
\begin{equation}
H_{lat}=\sum_{i=-\infty }^{\infty }(c_{i}^{\dagger }c_{i+1}+h.c.)+\frac{1}{2}%
\sum_{j=1}^{2}(u_{1}+(-1)^{j}u_{2})c_{j}^{\dagger }c_{j},  \notag
\end{equation}
at half-filling ($2k_{F}=\pi $) and with potential present at two adjacent
sites.

In Eq.\ (\ref{impHam}) we let $u_{1,2}(x)=u_{1,2}\delta _{a(1,2)}(x)$. Here $%
u_{1,2}$ are dimensionless quantities and $a_{1,2}$ is the range of the
corresponding regularization of the $\delta $-function; at the end the limit 
$a_{1,2}\rightarrow 0$ will be taken.

Combining the two chiral fermions into a spinor quantity $\Psi ^{\dagger
}=(\psi _{R}^{\dagger },\psi _{L}^{\dagger })$ we can write the equation of
motion as 
\begin{eqnarray}
i\partial _{t}\Psi  &=&v_{F}%
\begin{pmatrix}
i\partial _{x}+u_{1}(x) & u_{2}^{\ast }(x) \\ 
u_{2}(x) & -i\partial _{x}+u_{1}(x)%
\end{pmatrix}%
\Psi ,  \label{eq-motion} \\
&=&v_{F}(i\sigma _{3}\partial _{x}+u_{1}(x)+u_{2}^{\prime }(x)\sigma
_{1}+u_{2}^{\prime \prime }(x)\sigma _{2})\Psi ,  \notag
\end{eqnarray}%
with real and imaginary parts $u_{2}=u_{2}^{\prime }+iu_{2}^{\prime \prime }$. 
The formal solution of \ (\ref{eq-motion})\ is 
\begin{eqnarray}
\Psi (x,t) &=&e^{i\omega t}\mathcal{T}(x,a)\Psi (a,0),  \label{eigenstate} \\
\mathcal{T}(x,a) &=&\mathcal{T}_{x}\exp i\int_{a}^{x}{dy}\,\sigma
_{3}[v_{F}^{-1}\omega   \notag \\
&&+u_{1}(y)+u_{2}^{\prime }(y)\sigma _{1}+u_{2}^{\prime \prime }(y)\sigma
_{2}],
\end{eqnarray}%
where $\mathcal{T}_{x}$ stands for the $x$-ordering operator. In the long
wavelength limit, $\omega \rightarrow 0$, we obtain 
\begin{eqnarray}
\mathcal{T}(x,a) &=&\exp [i\omega v_{F}^{-1}\sigma _{3}(x-a)],\quad x<0,
\label{eq:defT0} \\
&=&\exp [i\omega v_{F}^{-1}\sigma _{3}x]\mathcal{T}_{0}\exp [-i\omega
v_{F}^{-1}\sigma _{3}a],\quad x>0,  \notag
\end{eqnarray}%
with $\mathcal{T}_{0}$ depending on $u_{1,2}(x)$. We assume first that the
regularized $\delta $-functions in the definitons of $u_{1,2}(x)$ are
identical, which means $a_{1}=a_{2}$. In this case 
\begin{equation}
\mathcal{T}_{0}=\exp (iu_{1}\sigma _{3}-u_{2}^{\prime }\sigma
_{2}+u_{2}^{\prime \prime }\sigma _{1})=\cosh b+\bm {b}\bm {\sigma}\frac{%
\sinh b}{b},  \label{eq:transfer-b}
\end{equation}%
with ${\bm b}=(u_{2}^{\prime \prime },-u_{2}^{\prime },iu_{1})$ and $%
b^{2}=|u_{2}|^{2}-u_{1}^{2}$. This transfer matrix simplifies in two
important cases : i) point-like impurity, when $u_{1}=|u_{2}|$ and ii)
purely backward scattering impurity, $u_{1}=0$, $u_{2}\neq 0$. We have 
\begin{eqnarray}
\mathcal{T}_{0} &=&%
\begin{pmatrix}
1+i|u_{2}| & iu_{2}^{\ast } \\ 
-iu_{2} & 1-i|u_{2}|%
\end{pmatrix}%
,\quad u_{1}=|u_{2}|, \\
\mathcal{T}_{0} &=&%
\begin{pmatrix}
\cosh |u_{2}| & ie^{-i\phi }\sinh |u_{2}| \\ 
-ie^{i\phi }\sinh |u_{2}| & \cosh |u_{2}|%
\end{pmatrix}%
,\quad u_{1}=0  \notag
\end{eqnarray}%
with $e^{i\phi }=u_{2}/|u_{2}|$.

Let us also consider two further limiting cases, when the range of the
forward scattering potential $u_{1}(x)$ is much longer (shorter) than the
range of the backward scattering potential $u_{2}(x)$, i.e.\ $a_{1}\gg a_{2}$
($a_{1}\ll a_{2}$ ). In these two cases we get 
\begin{equation}
\mathcal{T}_{0} =%
\begin{pmatrix}
e^{iu_{1}}\cosh |u_{2}| & ie^{-i\phi }\sinh |u_{2}| \\ 
-ie^{i\phi }\sinh |u_{2}| & e^{-iu_{1}}\cosh |u_{2}|%
\end{pmatrix}%
\end{equation}
 for $a_{1}\gg a_{2} $ and 
\begin{equation}
\mathcal{T}_{0} =\cos u_{1}%
\begin{pmatrix}
\cosh |u_{2}|+i\tan u_{1} & ie^{-i\phi }\sinh |u_{2}| \\ 
-ie^{i\phi }\sinh |u_{2}| & \cosh |u_{2}|-i\tan u_{1}%
\end{pmatrix}%
\end{equation}
 for $a_{1}\ll a_{2}$.  
By construction, the determinant of the transfer matrix is unity for
arbitrary $u_{1,2}(x)$.

The elements of the scattering matrix for the incident waves $\sim e^{ikx}$
and $\sim e^{-ikx}$ at $x<0$ and $x>0$, respectively, are defined by the
equation 
\begin{equation}
\begin{pmatrix}
t & \tilde{r} \\ 
0 & 1%
\end{pmatrix}%
=\mathcal{T}_{0}.%
\begin{pmatrix}
1 & 0 \\ 
r & \tilde{t}%
\end{pmatrix}
\label{defSmat}
\end{equation}%
Hence the scattering matrix is determined by the elements of the
transfer-matrix $\mathcal{T}$ as 
\begin{equation*}
\mathcal{S}\equiv 
\begin{pmatrix}
t & \tilde{r} \\ 
r & \tilde{t}%
\end{pmatrix}%
=\mathcal{T}_{22}^{-1}%
\begin{pmatrix}
1 & \mathcal{T}_{12} \\ 
-\mathcal{T}_{21} & 1%
\end{pmatrix}%
\end{equation*}%
In the above two cases with $a_{1}=a_{2}$ we have 
\begin{eqnarray}
\mathcal{S} &=&\frac{1}{1-i|u_{2}|}%
\begin{pmatrix}
1 & iu_{2}^{\ast } \\ 
iu_{2} & 1%
\end{pmatrix}%
,\quad u_{1}=|u_{2}|,  \label{Smat@u1=u2} \\
\mathcal{S} &=&%
\begin{pmatrix}
1/\cosh |u_{2}| & ie^{-i\phi }\tanh |u_{2}| \\ 
ie^{i\phi }\tanh |u_{2}| & 1/\cosh |u_{2}|%
\end{pmatrix}%
,\quad u_{1}=0.  \notag
\end{eqnarray}

\noindent Whereas for model potentials of different range we obtain 
\begin{equation}
\mathcal{S} =e^{iu_{1}}%
\begin{pmatrix}
1/\cosh |u_{2}| & ie^{-i\phi }\tanh |u_{2}| \\ 
ie^{i\phi }\tanh |u_{2}| & 1/\cosh |u_{2}|%
\end{pmatrix}%
 \label{Smat@a1>a2}
\end{equation}%
 for $a_{1}\gg a_{2}$  and 
\begin{equation}
\mathcal{S} =\frac{1}{\cosh |u_{2}|-i\tan u_{1}}%
\begin{pmatrix}
1/\cos u_{1} & ie^{-i\phi }\sinh |u_{2}| \\ 
ie^{i\phi }\sinh |u_{2}| & 1/\cos u_{1}%
\end{pmatrix} , 
 \label{Smat@a1<a2}
\end{equation}%
for $a_{1}\ll a_{2}$.  
We see that if the forward scattering $u_{1}$ has wider range, than $u_{2}$,
then Eq.\ (\ref{Smat@a1>a2}) differs from the second Eq.\ (\ref%
{Smat@u1=u2}) only by the overall phase which is unimportant in physical
observables. At the same time, if the model regularization of $u_{1}$ is
narrower than $u_{2}$, the resulting  Eq.\ (\ref{Smat@a1<a2}) closely
resembles the $S$-matrix in the double-barrier situation. \cite{Polyakov2003}

Expanding the reflection coefficient in powers of $u_{1,2}$ and using (\ref%
{eq:transfer-b}), (\ref{Smat@a1>a2}),  (\ref{Smat@a1<a2}) we have 
\begin{eqnarray}
|r| &\simeq &|u_{2}|-\frac{1}{3}|u_{2}|^{3},\quad a_{1}\gg a_{2},  \notag \\
|r| &\simeq &|u_{2}|-\frac{1}{6}|u_{2}|(u_{1}^{2}+2|u_{2}|^{2}),\quad
a_{1}=a_{2}, \\
|r| &\simeq &|u_{2}|-\frac{1}{6}|u_{2}|(3u_{1}^{2}+2|u_{2}|^{2}),\quad
a_{1}\ll a_{2},  \notag
\end{eqnarray}%
which shows that the forward scattering amplitude $u_{1}$ enters the
reflection coefficient in the third order, i.e. beyond the Born
approximation. Moreover the contribution of $u_{1}$ to the $S$-matrix
depends also on the details of the microscopic Hamiltonian and the
regularization of the model $\delta $-function potentials. This complication
happens at the step (\ref{eigenstate}); however, once the transfer matrix $%
\mathcal{T}_{0}$ in (\ref{eq:defT0}) is known, the connection to the
scattering state basis is straightforward.

Returning to (\ref{scat-states}), (\ref{ChiralScatStates}), we notice that $%
\psi _{1}(x)$ obeys the equation of motion for right movers, and the spinor $%
\Psi _{R}^{\dagger }=(\psi _{1}^{+}(x),r\psi _{1}^{+}(-x))$ is an eigenstate
(\ref{eigenstate}) of (\ref{eq-motion}) with the above property (\ref%
{defSmat}). We thus clarified the scattering states representation in the
framework of the chiral Hamiltonian (\ref{genHam}).

In the bosonization formulation, (\ref{bosonHam}), the assumption is made
that the forward scattering $u_{1}(x)$ can be completely removed from the
Hamiltonian. This statement is not obvious, in general. Consider the
redefinition $\varphi (x)=\tilde{\varphi}(x)-u_{1}sgn_{a1}(x)/2$, which
eliminates the term with $u_{1}$ from Eq.\ (\ref{bosonHam}), where the
regularized sign function $sgn_{a1}(x)=-1+2\int_{-\infty }^{x}dy\,\delta
_{a1}(y)$ is used. We have the backward scattering in the form 
\begin{eqnarray}
u_{2}(x)\cos (2\varphi (x)) &=&u_{2}(x)\cos (2\tilde{\varphi}%
(x)-u_{1}sgn_{a1}(x)),  \notag \\
&\rightarrow &u_{2}\,\delta (x)\cos (2\tilde{\varphi}(x)),  \notag \\
&\rightarrow &u_{2}\,\delta (x)\cos (2\tilde{\varphi}(x))\cos u_{1},
\end{eqnarray}%
for the above cases of regularization with $a_{1}\gg a_{2}$ and $a_{1}\ll
a_{2}$, respectively.

\section{Unitary transformation}

\label{sec:AppA}

We consider a transformation of the operators $J_{j}$ 
\begin{equation*}
O_{j}(x)=U^{\dagger }J_{j}(x)U
\end{equation*}%
with 
\begin{eqnarray}
U &=&\exp i\int_{-\infty }^{\infty }dx\,dy\,\theta (x-y)B_{j}(y)J_{j}(x), \\
&=&\exp i\int_{-\infty }^{\infty }dy\,B_{j}(y)\Theta _{j}(y), \\
\Phi _{j}(y) &=&\int_{y}^{\infty }dx\,J_{j}(x).
\end{eqnarray}%
In our case we have $B_{j}(x)=B_{j}\delta (x)$ and we may denote the
direction of $\mathbf{B}$ as "1", i.e.\ $\mathbf{B}\Vert \hat{e}_{1}$. Then
we have the following set of equalities from the Kac-Moody relations (\ref%
{KacMoody}) 
\begin{eqnarray}
\lbrack J_{1}(y),\Phi _{1}(x)] &=&\frac{i}{4\pi }\delta (x-y)  \notag \\
{}[J_{1}(y),i\int_{-\infty }^{\infty }dx\,B_{1}(x)\Theta _{1}(x)] &=&-\frac{1%
}{4\pi }B_{1}(y) \\
J_{1}(y)U &=&U(J_{1}(y)-\frac{1}{4\pi }B_{1}(y))  \notag
\end{eqnarray}%
And for $J^{+}(x)=J_{3}(x)+iJ_{2}(x)$ one has 
\begin{eqnarray}
\lbrack J^{+}(x),J_{1}(y)] &=&J^{+}(x)\delta (x-y)  \notag \\
{}[J^{+}(x),\Theta _{1}(y)] &=&J^{+}(x)\theta (x-y) \\
{}[J^{+}(x),i\int_{-\infty }^{\infty }dy\,B_{1}(y)\Theta _{1}(y)]
&=&iJ^{+}(x)\int_{-\infty }^{x}dy\,B_{1}(y)  \notag
\end{eqnarray}%
So that the relations between the operators are 
\begin{eqnarray}
O_{1}(x) &=&U^{\dagger }J_{1}(x)U=J_{1}(x)-\frac{1}{4\pi }B_{1}(x) \\
O_{3}(x)+iO_{2}(x) &=&U^{\dagger }J^{+}(x)U \\
&=&(J_{3}(x)+iJ_{2}(x))\exp i\int_{-\infty }^{x}dy\,B_{1}(y)  \notag
\end{eqnarray}%
The transformation $U$ shifts the value of $O_{1}(x)$ by a $c-$number, so
that strictly speaking, the operators $O_{j}(x)$ do not obey the same
Kac-Moody algebra in the region where $B_{1}(x)\neq 0$. It is clear, though
that this might affect only the physically non-observable component of the
fermionic density and is therefore not important.

\section{Equations of motion in current operator formalism}

\label{sec:AppC}

In this section we focus on the time evolution of the observables $J_3(-x)$, 
$\widetilde J_3(x)$, Eqs.\ (\ref{def:density}), (\ref{def:current}), using
the relation $i\partial_t A = [A, H]$. 
It is more convenient to use the representation (\ref{fictmagnfield}), with $%
O(x)=J_{3}(x)\Theta (-x)+\widetilde{J}_{3}(x)\Theta (x)$. In terms of
operators $O_{j}(x)$ one has 
\begin{eqnarray}
\rho _{s}(x) &=&(O_{3}(x)-O_{3}(-x)) \\
j_{s}(x) &=&v_{F}(O_{3}(x)+O_{3}(-x))  \notag
\end{eqnarray}%
Using (\ref{KacMoody}) and formulas of Appendix \ref{sec:AppA}, it can be
easily shown that we have $-\partial _{t}O_{3}(x)=\partial _{x}O_{3}(x)$ in
the leads, $|x|>L$, whereas in the interacting region $a<|x|<L$ we obtain 
\begin{equation}
-\partial _{t}O_{3}(x)=\partial _{x}O_{3}(x)-g\partial _{x}O_{3}(-x),
\label{eq:motion}
\end{equation}%
where we let $v_{F}=1$. Since $O_{3}(x)$ is the \textit{observable} density,
the relation (\ref{eq:motion}) looks deceptively close to the desired
answer. However, in the region of the barrier $|x|<a$ (with $\mathbf{B}\Vert
O_{1}$) we have 
\begin{equation*}
-\partial _{t}O_{3}(x)=\partial _{x}O_{3}(x)+B(x)O_{2}(x),
\end{equation*}%
and the $O_{2}$ component comes into play. (In other words, for the
partially reflecting barrier, the observable of our problem, $\widetilde{J}%
_{3}(x)$, includes the \textit{eigenmode} operator $J_{2}(x)$.) Therefore it
is necessary to consider also the time evolution of operator $O_{2}$, which
reads 
\begin{equation}
-\partial _{t}O_{2}(x)=\partial _{x}O_{2}(x)-4\pi gO_{1}(x)O_{3}(-x),
\label{eq:motion2}
\end{equation}%
at $a<|x|<L$ and $-\partial _{t}O_{2}(x)=\partial _{x}O_{2}(x)+B(x)O_{3}(x)$%
, at $|x|<a$. The four-fermion object $O_{1}(x)O_{3}(-x)$ is not reduced to $%
O_{2}(-x)$ and therefore we do not obtain a closure of the equations of
motion.

Our situation is somewhat similar to the well-known Heisenberg model of
spins with pairwise exchange interaction. Due to the non-commutative
character of spin operators, the attempt to write the exact time evolution
of spin leads to an infinite series of coupled equations, including ever
increasing numbers of spins. The approximate solutions are obtained by
truncating this series at some step, according to a certain criterion, and
discarding higher terms as inessential. \cite{Izyumov1988} In magnetism,
these criteria might be (i) low temperatures in the magnetically ordered
phase, $T\ll T_c$, (ii) large value of spin $S$ or (iii) wide range of
interaction. In our case, we choose the criterion of relevancy of operators,
i.e.\ their importance in the logarithmically divergent theory.

We now briefly discuss the dynamics of $O_{4}(x)\equiv 4\pi O_{1}(x)O_{3}(-x)$. 
The compound operator $O_{4}(x) $ consists of four fermion
operators and may be decomposed into the normally ordered part and a part,
representing internal contractions. To clarify the notation we write $\psi
_{1}(x)=a_{\uparrow }(x)$, $\psi _{2}(-x)=a_{\downarrow }(x)$, then we have 
\begin{eqnarray}
O_{4}(x) &=&\pi (a_{\uparrow }^{\dagger }(x)a_{\downarrow }(x)+a_{\downarrow
}^{\dagger }(x)a_{\uparrow }(x))  \notag \\
&&\times (a_{\uparrow }^{\dagger }(-x)a_{\uparrow }(-x)-a_{\downarrow
}^{\dagger }(-x)a_{\downarrow }(-x))
\end{eqnarray}%
Contracting here, we get $O_{4}(x)-\!:\!O_{4}(x)\!\!:=-x^{-1}O_{2}(\bar{x})$
with (cf.\ Eq.\ (\ref{preciseOPE})) 
\begin{equation}
O_{2}(\bar{x})=\frac{1}{4i}(a_{\uparrow }^{\dagger }(x)a_{\downarrow
}(-x)+a_{\uparrow }^{\dagger }(-x)a_{\downarrow }(x))+h.c.
\end{equation}%
which does not coincide with $O_{2}(x)=(2i)^{-1}a_{\uparrow }^{\dagger
}(x)a_{\downarrow }(x)+h.c.$ We may write 
\begin{equation}
O_{2}(\bar{x})\simeq O_{2}(-x)-x\partial _{x}O_{2}(-x)+\ldots 
\end{equation}%
and, dropping the less relevant higher derivatives, obtain from (\ref%
{eq:motion2}) 
\begin{equation}
-\partial _{t}O_{2}(x)\simeq \partial _{x}O_{2}(x)+g(x^{-1}-\partial
_{x})O_{2}(-x)-g:\!O_{4}(x)\!:  \notag
\end{equation}%
%     \begin{equation}               (\partial_t  + \partial_x) O_2(x) \simeq 
%         g ( - x^{-1} + \partial_x) O_2(-x)   + g :\!O_4(x)\!:      \end{equation}
which means that  the most relevant part of the  operator $O_{4}(x)$ has the form 
\begin{equation}
O_{4}(x)\simeq (-1/x+\partial _{x})O_{2}(-x).  \label{approxO4asO2}
\end{equation}

Regarding the normally ordered part $:O_{4}(x):$, its time evolution is
given by 
\begin{eqnarray*}
-\partial _{t}:\!O_{4}(x)\!: &=&4\pi :\!O_{3}(-x)\partial
_{x}O_{1}(x)-O_{1}(x)\partial _{x}O_{3}(-x)\!: \\
&&+4\pi g:\!O_{5}(x)\!: \\
O_{5}(x) &=&4\pi O_{1}(x)O_{3}^{2}(-x)+O_{1}(x)\partial _{x}O_{3}(x)
\end{eqnarray*}%
i.e.\ described by new operators, not reduced to simple spatial derivatives
of $:O_{4}(x):$. Such derivative objects (i) obey an infinite set of coupled
equations and (ii) possess formal scaling dimension greater than two, and
hence are irrelevant in the RG sense.

Our strategy of such truncation (\ref{approxO4asO2}) is 
successful only in part, as we explain in the next 
Appendix \ref{sec:AppC2}.

\section{Solution of the equation of motion}
\label{sec:AppC2}

Taking a Fourier transform in time $t$, and introducing the vector 
\begin{equation*}
\Psi(x,\omega) = \left[ O_3(x,\omega) , O_3(-x,\omega), O_2(x,\omega) ,
O_2(-x,\omega) \right]
\end{equation*}
the above truncated set of equations can be written as 
% \begin{eqnarray}
\begin{align}
\partial_x \Psi(x,\omega) &= \hat D_0 (\omega, x) \vert_{g=0}
\Psi(x,\omega) , \quad |x| > L  \label{equ:mot_trunc} \\
\hat D_1 \partial_x \Psi(x,\omega) &= \hat D_0 (\omega, x) \Psi(x,\omega) ,
\quad a< |x| < L  \notag 
\end{align} % \end{eqnarray}
with 
\begin{eqnarray}
\hat D_0 (\omega, x) &=& 
\begin{pmatrix}
i\omega , & 0 , & 0, & 0 \\ 
0 , & -i\omega , & 0, & 0 \\ 
0, & 0, & i\omega , & -g f(x) \\ 
0, & 0, & -g f(x) , & -i\omega%
\end{pmatrix}
\\
\hat D_1 &=& 
\begin{pmatrix}
1 , & - g , & 0, & 0 \\ 
- g , & 1 , & 0, & 0 \\ 
0, & 0, & 1 , & -g \\ 
0, & 0, & -g , & 1%
\end{pmatrix}%
\end{eqnarray}
and the fermionic correlation function at finite temperature 
\begin{eqnarray}
f(x) = \frac{2\pi T}{\sinh 2\pi T x}
\end{eqnarray}
The solution to (\ref{equ:mot_trunc}) is given by the $x-$ordered exponent 
\begin{eqnarray}
\Psi(x,\omega) &=& \hat U (x,y) \Psi(y,\omega) \\
\hat U (x,y)&=& T_x \exp \int_y^x dz\,\hat D_1^{-1} \hat D_0 (\omega, z) ,
\label{T-ordered}
\end{eqnarray}
similarly to (\ref{eigenstate}), but now with transfer-matrix $\hat U (x,y)$
for densities. The transfer-matrix in the region of the barrier $|x|<a$ is
already known and given by 
\begin{equation}
\hat U (a,-a) = 
\begin{pmatrix}
\cos 2\theta , & 0, & \sin 2\theta , & 0 \\ 
0 , & \cos 2\theta , & 0, & -\sin 2\theta \\ 
- \sin 2\theta , & 0, & \cos 2\theta , & 0 \\ 
0, & \sin 2\theta , & 0 , & \cos 2\theta%
\end{pmatrix}%
\end{equation}
The overall transfer-matrix over the interaction region is 
\begin{equation}
\hat U (L,-L) = \hat U (L,a). \hat U (a, -a). \hat U ( -a, -L)
\end{equation}
Finally we observe that the definition of $\Psi(x)$ implies that the
transfer matrix connecting opposite points should satisfy the boundary
condition: 
\begin{equation}
\hat U (x,-x)\Psi(-x,\omega) = \hat I .\Psi(-x,\omega)  \label{bound-cond}
\end{equation}
\begin{equation}
\hat I = 
\begin{pmatrix}
0, & 1, & 0, & 0 \\ 
1, & 0 , & 0, & 0 \\ 
0, & 0, & 0 , & 1 \\ 
0, & 0, & 1 , & 0%
\end{pmatrix}%
\end{equation}
and we should choose a solution, by demanding that the scattered component, $%
O_2$, is absent in the incoming wave, i.e.\ $\Psi_3(-\infty,\omega) =0 $.

\begin{widetext}
We consider here only the technically feasible limit, $\omega =0$. In this case 
$\hat U (x , L)  = \hat U (- L,-x ) = 1 $  in the leads, $x>L$. 
In the interacting region
     \begin{eqnarray}
    \hat U (L,a) &=&
     \begin{pmatrix}
     1, & 0,&0,& 0 \\
      0,& 1 , & 0,& 0 \\
      0,& 0,&  e^{-g^2 \Lambda}\cosh g\Lambda ,
      &-e^{-g^2 \Lambda} \sinh g\Lambda \\
      0,&  0,& -e^{-g^2 \Lambda} \sinh g\Lambda ,
      &e^{-g^2 \Lambda} \cosh g\Lambda
     \end{pmatrix}
     \label{transfer-g-static}\\
     \Lambda &=& (1-g^2)^{-1}\int_{a}^L dx\, f(x)
%     \\  &=&
=  (1-g^2)^{-1} \ln\left(  \frac{\tanh \pi T L}{\tanh \pi T a}\right) 
     \label{def:LambdaEqMo}
    \end{eqnarray}
A similar contribution on the negative semiaxis, 
$ \hat U ( -a, -L) $,  is obtained from  (\ref{transfer-g-static}) 
by the change $\Lambda\to -\Lambda$. 
Overall we have 
  \[ \hat U (\infty,-\infty) =
     \begin{pmatrix}
     \cos 2\theta, & 0,&
     - e^{g^2 \Lambda}\sin 2\theta \cosh g\Lambda,
     &-e^{g^2 \Lambda}\sin 2\theta \sinh g\Lambda \\
      0 ,& \cos 2\theta ,
      & e^{g^2 \Lambda} \sin 2\theta \sinh g\Lambda,
      &  e^{g^2 \Lambda} \sin 2\theta \cosh g\Lambda \\
       e^{-g^2 \Lambda}\sin 2\theta \cosh g\Lambda,
      & e^{-g^2 \Lambda}\sin 2\theta \sinh g\Lambda,
      &  \cos 2\theta , &0 \\
      -e^{-g^2 \Lambda}\sin 2\theta \sinh g\Lambda,
      & -e^{-g^2 \Lambda} \sin 2\theta \cosh g\Lambda,
      &  0 ,& \cos 2\theta
     \end{pmatrix}
    \]
Solving   Eq.\ (\ref{bound-cond}) together with the above condition
$\Psi_3(-\infty) =0 $,  we find
    \begin{equation}
    \Psi(-\infty) = ( \cos^2 \theta + e^{2g\Lambda}\sin^2 \theta ,
    \cos^2 \theta  - e^{2g\Lambda}\sin^2 \theta , 0 ,- e^{g(1-g)\Lambda}\sin 2\theta)
    \label{equmo-solu}
    \end{equation}
which is the desired solution. Let us briefly discuss it. 

\end{widetext}

We observe that the d.c.\ current $j(x) = O_3(x) + O_3(-x)$ is independent
of $x$, as it should be. This property follows from the form of matrix (\ref%
{transfer-g-static}) and condition (\ref{bound-cond}). The relation of the
current to the voltage bias defines the conductance of the system. In our
definitions, the voltage bias is proportional to the difference in number of
incoming left- and right-going electrons, which is $2O_3(-\infty) =
2\Psi_1(-\infty)$, so that the conductance is 
\begin{eqnarray}
G &=& \frac{j(x)}{2O_3(-\infty)} = \frac12 \left(\frac {O_3(\infty)}{
O_3(-\infty)} +1 \right)  \notag \\
&=& \frac{ \cos^2 \theta}{\cos^2 \theta + e^{2g\Lambda}\sin^2 \theta}
\end{eqnarray}
which coincides with the above result (\ref{soluRG1st}).

At first glance, the result (\ref{equmo-solu}) should thus be viewed as
satisfactory. However there are details, involved in its derivation, which
spoil the significance of this approach. First we notice that, by keeping
less relevant gradient term in the above relation, (\ref{approxO4asO2}), we
get the appearance of $g^2$ terms in the definition of $\Lambda$, eq.\ (\ref%
{def:LambdaEqMo}), and also in the exponents, $e^{g^2\Lambda}$, in (\ref%
{transfer-g-static}) -- (\ref{equmo-solu}). Further, trying a different
approximation instead of (\ref{approxO4asO2}), e.g. $O_4(x) \simeq (-1/x +
\partial_x) O_2(x) $, would not produce the above renormalized value of
conductance.

Therefore we conclude that using the equations of motion method and
persisting with the current operator algebra leads to ambiguities, connected
with the non-local character of the interaction. 

\section{Calculation of corrections: zero temperature}

\label{sec:AppD}

Proceeding as described in Sec.\ \ref{sec:CorrCondA}, 
we find corrections to the conductance, or to 
$Y$ in (\ref{def:Y}). In the first order of $g$ the correction from the
diagrams of the type $\{4\}$ has the form 
\begin{equation}
\delta Y^{\{4\}}=-g(1-Y^{2})\int_{a}^{L}\frac{dx}{x}=-g(1-Y^{2})\Lambda _{0}
\label{corr(4)}
\end{equation}

In the second order of $g$ we have two sets of diagrams for corrections to
the conductance. The diagrams containing two fermionic loops lead to the
following expression : 
\begin{eqnarray}
\delta Y^{\{4,2\}} &=&g^{2}Y(1-Y^{2})\int_{a}^{L}\frac{dx_{1}dx_{2}}{%
2(x_{1}+x_{2})^{2}}  \notag \\
&=&\frac{1}{2}g^{2}Y(1-Y^{2})(\Lambda _{0}-2\ln 2)
\end{eqnarray}%
where $x_{1,2}$ are two points of fermionic interaction.

All diagrams consisting of only one loop, $\{6\}$ in above notation, produce
the correction 
\begin{equation}
\delta Y^{\{6\}} = -g^2Y(1-Y^2) \int_a^L \frac{dx_1 dx_2}{x_1x_2} =
-g^2Y(1-Y^2)\Lambda_0^2  \label{corr(6)}
\end{equation}

We should mention that individual diagrams in the set $\{6\}$ (and higher
orders) may contain singularities of the form $(x_1-x_2)^{-1}$ etc. However
these singularities cancel each other in the resulting expressions. For
example, one finds two particular contributions $x_1^{-1} (x_1-x_2)^{-1}$
and $x_2^{-1} (x_2-x_1)^{-1}$ whose sum leads to the above ``regular'' form, 
$(x_1x_2)^{-1}$, in (\ref{corr(6)}).

In the third order, we have to analyze the following types of the diagrams : 
$\{4,2,2\}$, $\{4,4\}$, $\{6,2\}$, $\{8\}$. We assume the ordered sequence, $%
L>x_1>x_2>x_3>a$, in all integrals below. Performing rather long computer
calculations (the stable routine in \textit{Mathematica} requires about ten
hours of computation time for dual-core 3 GHz processor), we obtain 
\begin{eqnarray}
\delta Y^{\{4,2,2\}} &=& -g^3(1-Y^2) \int{\prod dx_i}  \notag \\
&\times& \left[ \frac{3 Y^2} {(x_1+x_2+x_3)^3} + \frac{1} {(x_1+x_2-x_3)^3} %
\right]  \notag \\
&=& -g^3(1-Y^2)(1+Y^2)\frac {\Lambda_0} 4 ,  \label{corr(4,2,2)}
\end{eqnarray}%
\begin{eqnarray}
\delta Y^{\{4,4\}} &=& -g^3(1-Y^2)^2 \int{\prod dx_i}  \notag \\
&\times& \left[ \frac{1} {x_1(x_2+x_3)(x_1+x_2+x_3)} +(\mbox{symm.}) \right]
\notag \\
&=& -g^3(1-Y^2)^2\frac 14 (\Lambda_0^2 + 2\Lambda_0(1-\ln2) ) ,
\label{corr(4,4)}
\end{eqnarray}%
\begin{eqnarray}
\delta Y^{\{6,2\}} &=& 2g^3(1-Y^2)Y^2 \int{\prod dx_i}  \notag \\
&\times& \left[ \frac{1} {x_1(x_2+x_3)^2} +(\mbox{symm.}) \right]  \notag \\
&=& g^3(1-Y^2)Y^2 \Lambda_0(\Lambda_0 -2\ln 2) ,  \label{corr(6,2)}
\end{eqnarray}%
\begin{eqnarray}
\delta Y^{\{8\}} &=& 2g^3(1-Y^2) \int{\prod dx_i} \left[ \frac{1-3Y^2} {x_1
x_2 x_3} \right.  \notag \\
&+& \left. \frac{1-Y^2} {(x_1+x_2)(x_2+x_3)(x_3+x_1)} \right]
\label{corr(8)} \\
&=& g^3(1-Y^2)\left[\frac{1 - 3Y^2}3 \Lambda_0^3 + (1-Y^2)\frac{\pi^2}{12}%
\Lambda_0 \right].  \notag
\end{eqnarray}
In the final expressions (\ref{corr(4,2,2)}), (\ref{corr(4,4)}), (\ref%
{corr(8)}) we omitted finite parts, $O(\Lambda^0)$, which are unimportant in
the order $g^3$. These expressions lead to Eq.\ (\ref{eq:corY}) above.

\section{Calculation of corrections: finite temperature}

\label{sec:AppE}

In case of finite temperature we proceed in the same way as for $T=0$ , only
with the integration replaced by a summation over Matsubara frequencies.
Introducing the scaling variable $\Lambda $ according to (\ref{def:Lambda})
we have for $a\ll \xi =v_{F}/\pi T\ll L$ 
\begin{eqnarray}
\delta Y^{\{4\}} &=&-g(1-Y^{2})\Lambda ,  \notag \\
\delta Y^{\{4,2\}} &=&g^{2}Y(1-Y^{2})\int_{a/\xi }\frac{2d\bar{x}_{1}d\bar{x}%
_{2}}{\sinh ^{2}2(\bar{x}_{1}+\bar{x}_{2})}  \label{corr(4,2)} \\
&=&\frac{1}{2}g^{2}Y(1-Y^{2})(\Lambda -3\ln 2),  \notag
\end{eqnarray}%
where $x_{1,2}$ are two points of fermionic interaction, and $\bar{x}%
_{i}=x_{i}/\xi $. In the last equality in the right-hand side of Eq.\ (\ref%
{corr(4,2)}), we shall neglect the terms $\sim O(1)$.

Instead of $\Lambda _{0}^{2}$ in the expression (\ref{corr(6)}) for $\delta
Y^{\{6\}}$ we now have the following integral 
\begin{eqnarray}
&&\int_{a/\xi }^{L/\xi }{d\bar{x}_{1}d\bar{x}_{2}}\frac{3\cosh (\bar{x}_{1}+%
\bar{x}_{2})+\cosh (\bar{x}_{1}-\bar{x}_{2})}{\sinh (2\bar{x}_{1})\sinh (2%
\bar{x}_{2})\cosh (\bar{x}_{1}+\bar{x}_{2})}  \notag \\
&=&\int_{\tau _{0}}^{\tau _{L}}\frac{d\tau _{1}d\tau _{2}}{\tau _{1}\tau _{2}%
}\left( 1-\frac{1}{2}\frac{\tau _{1}\tau _{2}}{1+\tau _{1}\tau _{2}}\right) 
\label{TempCorr(6)} \\
&=&\Lambda ^{2}-F[\tau _{L},\tau _{0}]  \notag \\
&\rightarrow &\Lambda ^{2}-\frac{\pi ^{2}}{24},\quad T\gg v_{F}/L  \notag
\end{eqnarray}%
with $\tau _{i}=\tanh \bar{x}_{i}$ . The integration limits in (\ref%
{TempCorr(6)}) are $\tau _{0}=\tanh (a/\xi )$ and $\tau _{L}=\tanh (L/\xi )$%
. Here $F[\tau _{L},\tau _{0}]=L_{2}(-\tau _{0}\tau _{L})-(L_{2}(-\tau
_{L}^{2})+L_{2}(-\tau _{0}^{2}))/2$ and $L_{2}(x)=\sum_{1}^{\infty
}x^{n}/n^{2}$ is the dilogarithm function. We have $F[\tau _{L},\tau _{0}]=0$
in the limit $T\ll v_{F}/L$.

In contrast to the $T=0$ case, the expressions now become extremely
cumbersome in the third order of $g$. The summation over Matsubara
frequencies lead to expressions involving sums and products of exponents $%
\exp (\bar{x}_{i})$. A general criterion for the correctness of intermediate
expressions is the property a change of sign of each individual diagram upon
performing the  reflection transformation $z_{i}\rightarrow -z_{i}$.

An additional check of the validity of calculation is the observation that
we should get zero correction to $G=1$ in the absence of the barrier, $%
\theta =0$, and to $G=0$ for the fully reflecting barrier, $\theta =\pi /2$,
for arbitrary $T$. This is not a trivial statement, because each diagram in
third order contains a prefactor, either $\cos 8\theta $, or $\cos 4\theta $
or $\cos 0\theta =1$. We verified that the algebraic expressions canceled
each other in this case.

The actual expressions are not shown here, and we describe our method of 
analysis below.

Let us consider an expression of the form of a triple integral, relevant to
our discussion in Sec.\ \ref{sec:CorrCond}

\begin{equation}
f(a)=\int_{a}^{L}dz_{1}\int_{z_{1}}^{L}dz_{2}\int_{z_{2}}^{L}dz_{3}\bar{f}%
[z_{1},z_{2},z_{3}].
\end{equation}%
We assume that $f(a)\simeq -c\ln a+\ldots $ at $a\rightarrow 0$ and wish to
determine the coefficient $c$ in this asymptotic behavior. We have 
\begin{equation}
c=-\lim_{a\rightarrow 0}\frac{df(a)}{d\ln a}=\lim_{a\rightarrow
0}\int_{a}^{L}dz_{2}\int_{z_{2}}^{L}dz_{3}\,a\bar{f}[a,z_{2},z_{3}]
\end{equation}%
Here taking the limit in the integrand may lead to incorrect results. We
have two possibilities for the behavior of the quantity $\bar{f}%
_{2}[a,z_{2}]=\int_{z_{2}}^{L}dz_{3}\,a\bar{f}[a,z_{2},z_{3}]$, for small $a$%
, namely $\bar{f}_{2}[a\rightarrow 0,z_{2}]\sim a/(z_{2}+a)^{2}$ and $\bar{f}%
_{2}[a\rightarrow 0,z_{2}]\sim 1$. The first type of contribution, $\bar{f}%
_{2}\sim a/(z_{2}+a)^{2}$, is always important and survives in the limit of $%
T\rightarrow 0$. Recalling that $\bar{f}[a,z_{2},z_{3}]$ is exponentially
suppressed at $z_{i}\agt\xi $, we see that the case $\bar{f}%
_{2}[a,z_{2}]\sim 1$ contributes to  $c$ only at finite temperatures, $L\gg
\xi $ ; typically, we find $\bar{f}_{2}[a\rightarrow 0,z_{2}]\sim (\xi \cosh
(2z_{2}/\xi ))^{-1}$.

Therefore the recipe for the determination of  $c$ is first to determine the
limiting form of $\bar{f}_{2}$, taking $z_{2}\sim a\ll \xi $ ; in this case
the expressions are drastically simplified. This will determine $c$ at $%
T\rightarrow 0$. To this term we should add the contribution $%
\lim_{a\rightarrow 0}a\bar{f}[a,z_{2},z_{3}]$, integrated over $z_{2}$ in
the interval $(0,\infty )$ ; the sum of these two contributions gives the
value of $c$ at finite temperatures, $L\gg \xi $.

\section{"Non-universality of higher terms" in $\protect\beta-$function}

\label{sec:NU}

In a very general sense the $\beta $-function of the RG-equation is
non-universal: it depends on the precise definition of the scaling quantity 
$g$ .  Let the RG equation read 
\begin{equation}
dg_{r}/d\Lambda =\beta
(g_{r})=b_{2}g_{r}^{2}+b_{3}g_{r}^{3}+b_{4}g_{r}^{4}+\ldots   \label{NU:eq0}
\end{equation}%
and consider a change of variable $g_{r}=\bar{g}+c_{2}\bar{g}^{2}+c_{3}\bar{g%
}^{3}$. With the same accuracy we have 
\begin{eqnarray}
{d\bar{g}}/{d\Lambda } &=&\beta (\bar{g})=\beta (g)/(dg/d\bar{g})=b_{2}\bar{g%
}^{2}+b_{3}\bar{g}^{3}  \notag \\
&&+(b_{4}+b_{3}c_{2}+b_{2}(c_{2}^{2}-c_{3}))\bar{g}^{4}+\ldots 
\label{NU:eq1}
\end{eqnarray}%
which shows that only the two first two coefficients $b_{2}$ and $b_{3}$ are
invariant. Since $b_{4}$ is associated with the three-loop contribution, one
may speak about a non-universality of the$\ \beta $-function beyond two-loop
order.

Let us elucidate the origin of the Eq.\ (\ref{NU:eq1}), and compare it with
the equation obtained in the Callan-Symanzik approach. In the Gell-Mann--Low
approach we perform our calculations with the running coupling constant $%
g_{r}$ and add counterterms to the Hamiltonian in order to achieve the
finite value of the bare constant $g_{0}$. One can show that the above value
of $dg_{r}/d\Lambda =-(\partial g_{0}/\partial \Lambda )/(\partial
g_{0}/\partial g_{r})$ is obtained from the expression 
\begin{eqnarray}
g_{0} &=&g_{r}-b_{2}\Lambda g_{r}^{2}-(b_{3}\Lambda -b_{2}^{2}\Lambda
^{2})g_{r}^{3}  \notag \\
&&-(b_{4}\Lambda -\frac{5}{2}b_{2}b_{3}\Lambda ^{2}+b_{2}^{3}\Lambda
^{3})g_{r}^{4}+\ldots   \label{NU:eq2}
\end{eqnarray}%
If we make the above change $g_{r}\rightarrow \bar{g}$ in the right-hand
side of (\ref{NU:eq2}) and use the same recipe for $d\bar{g}/d\Lambda $, we
return to (\ref{NU:eq1}).

In the CS approach we fix the bare value $g_{0}$ and calculate the
renormalized quantity $g_{r}(g_{0},\Lambda )$. Inverting the equation $%
g_{0}(g_{r},\Lambda )$ and using the same recipe we find $dg_{r}/d\Lambda $.
One can check that 
\begin{eqnarray}
g_{r} &=&g_{0}+b_{2}\Lambda g_{0}^{2}+(b_{3}\Lambda +b_{2}^{2}\Lambda
^{2})g_{0}^{3}  \notag \\
&&+(b_{4}\Lambda +\frac{5}{2}b_{2}b_{3}\Lambda ^{2}+b_{2}^{3}\Lambda
^{3})g_{0}^{4}+\ldots   \label{NU:eq3}
\end{eqnarray}%
Assume now that we have changed the starting value $g_{0}=\bar{g}_{0}+c_{2}%
\bar{g}_{0}^{2}+c_{3}\bar{g}_{0}^{3}+\ldots $, according to above
prescription. Evidently, the expansion for $g_{r}(\bar{g}_{0},\Lambda )$
will be different from (\ref{NU:eq3}), as well as the inverse function $\bar{%
g}_{0}(g_{r},\Lambda )$ will not coincide with (\ref{NU:eq2}). One can
verify however that the $\beta $-function, $-(\partial \bar{g}_{0}/\partial
\Lambda )/(\partial \bar{g}_{0}/\partial g_{r})$, is given by the same Eq.\ (%
\ref{NU:eq0}) and not by Eq.\ (\ref{NU:eq1}). This fact becomes rather
obvious, when we notice that only the left-hand side of (\ref{NU:eq2}) is
changed and it corresponds to a change in the boundary condition for $%
g_{r}(\Lambda )$ rather than to a change in the differential equation for it.

%%%%%%%%%%%%%%%%%%%%%%%%%%%%%%%%%%%%%%%%%

\section{Comparison to CFT solution}

\label{sec:CFT}

Consider the differential equation 
\begin{equation}
\lbrack -\partial _{y}^{2}+se^{Ky}+e^{y}]\Psi (y)=0
\end{equation}%
such that $\Psi (y\rightarrow \infty )=0$ and $\Psi (y\rightarrow -\infty
)=y-y_{0}$. Here 
\begin{equation*}
s=K^{2}\exp (2(1-K)\Lambda )
\end{equation*}%
and $\Lambda =\ln (T_{0}/T)$. For given $s$ one finds a solution $\Psi (y)$
characterized by a unique value $y_{0}$. Then, according to Lukyanov, \cite%
{Lukyanov2007} the conductance can be determined as 
\begin{equation}
G(T)=1+Ks\frac{\partial y_{0}}{\partial s}=1+\frac{K}{2(1-K)}\frac{\partial
y_{0}}{\partial \Lambda }
\end{equation}

We now fix the overall temperature scale $\tau =T/T_{0}$ by the high-$T$
behavior, $G\simeq 1-\tau ^{2(K-1)}$, and ask what is the coefficient $C$ in
the asymptotic low temperature expression $G\simeq C\tau ^{2(K^{-1}-1)}$. To
answer this question, we slightly rewrite the differential equation.
Introducing a new variable $z=2e^{y/2}$ we have 
\begin{equation}
\lbrack -\partial _{z}^{2}-z^{-1}\partial _{z}+1+s(z/2)^{2(K-1)}]\Psi (z)=0
\label{eq:difeqmod}
\end{equation}%
At $s=0$ (i.e.\ $T\rightarrow \infty $) the solution is a modified Bessel
function $\Psi (z)=-2K_{0}(z)$. At infinitesimal $s$ we seek a solution in
perturbation theory as $\Psi (z)=-2K_{0}(z)+sR(z)+O(s^{2})$. After some
calculation we find the conductance $G(T)\simeq 1-Ks\mathcal{F}(K)$ with $%
\mathcal{F}(K)\equiv R(0)$  given by 
\begin{eqnarray}
\mathcal{F}(K) &=&\Gamma ^{2}[K]\int_{0}^{\infty }{}_{2}F_{1}(K,K,1;-\kappa )%
\frac{d\kappa }{1+\kappa }  \notag \\
&=&{\Gamma ^{4}[K]}/{\Gamma \lbrack 2K]}\;.
\end{eqnarray}%
In the opposite limit, $s\rightarrow \infty $ ($T=0$), we choose a variable $%
z=(2\sqrt{s}/K)e^{Ky/2}$ and arrive at an equation 
\begin{equation}
\lbrack -\partial _{z}^{2}-z^{-1}\partial
_{z}+1+s_{1}(z/2)^{2(K^{-1}-1)}]\Psi (z)=0,
\end{equation}%
with $s_{1}=s^{-1/K}K^{2(K^{-1}-1)}\ll 1$. The solution is found similarly
to (\ref{eq:difeqmod}) and we get $G(T)\simeq -s\partial _{s}s_{1}\mathcal{F}%
(K^{-1})$. Comparing the two solutions we obtain the prefactor 
\begin{equation}
C=\left( K^{3}\mathcal{F}(K)\right) ^{1/K}K^{-3}\mathcal{F}(K^{-1})
\label{prefac-Luk}
\end{equation}%
In the particular case of $K=1/2$ we have $C=\pi ^{4}/48=2.029\ldots $, in
accordance with Ref.\ [\onlinecite{Weiss1995}]. For $K=1/3$ we get $%
C=10.064558\ldots $, a value slightly different from $C=10.0638$ reported in 
\cite{Fendley1995,Fendley1995a}, where it was obtained by numerical solution
of an integral equation. If we take $c_{3}=1/4$ in (\ref{eq:matching}), then
we would obtain $C=2$ ($C=9$) for the cases of $K=1/2$ ($K=1/3$), in good
agreement with the above CFT values. It should be noted, however, that apart
from this agreement in the matching coefficient $C$, the theory in \cite%
{Lukyanov2007} provides the value of unity for $G$ in the clean wire,
whereas the previous works \cite{Weiss1995,Fendley1995,Fendley1995a} gave $%
G=K$ in that case.

%\bibliography{../../DScThesis/Russian/thesis}

\end{document}